\documentclass[preprint,preprintnumbers,amsmath,amssymb,floatfix,nofootinbib,superscriptaddress]{revtex4}

\usepackage{graphicx}
\usepackage{caption}
\usepackage{slashed}
\usepackage{xcolor}
\usepackage{hyperref}
\hypersetup{
    colorlinks=true,
    linkcolor=blue,
    citecolor=blue,
    filecolor=magenta,      
    urlcolor=blue,
    }

\begin{document}

\title{NLO electroweak and QCD corrections to the production of a photon with three charged lepton plus missing energy at the LHC}

\author{Huanfeng Cheng}
\email{huanfeng@buffalo.edu}
\affiliation{Department of Physics, University at Buffalo,
The State University of New York, Buffalo, NY 14260-1500, U.S.A.}
\author{Doreen Wackeroth}
\email{dw24@buffalo.edu}
\affiliation{Department of Physics, University at Buffalo,
The State University of New York, Buffalo, NY 14260-1500, U.S.A.}

\date{\today}
\begin{abstract}
Electroweak (EW) triboson production processes with at least one heavy gauge boson are of increasing interest at the Large Hadron Collider (LHC) as direct precision probes of one of the least-tested sectors of the Standard Model (SM), the quartic couplings of the EW gauge bosons. These processes therefore offer promising opportunities for searches for indirect signals of Beyond-the-SM (BSM) physics. In this paper, we present results for fiducial cross sections at next-to-leading-order (NLO) EW and NLO QCD to $p\;p\to e^{+}\;\nu_{e}\;\mu^{+}\;\mu^{-}\;\gamma$ at the 13 TeV
LHC. This signature includes the triboson production process $p\;p \to W^{+}\;Z\;\gamma$ with leptonic decays, $W^{+} \to e^{+}\;\nu_{e}$ and $Z \to \mu^{+}\;\mu^{-}$. The computation is based on the complete set of
leading-order (LO) contributions of $\mathcal{O}(\alpha^5)$ and on NLO EW and NLO QCD cross sections of $\mathcal{O}(\alpha^6)$ and $\mathcal{O}(\alpha^5 \alpha_\text{s})$, respectively, and thus off-shell effects, spin correlations and non-resonance contributions are fully taken into account. We construct a Monte Carlo framework which provides total and differential cross sections for a chosen set of basic analysis cuts. We find that while NLO EW corrections enhance the fiducial LO total cross section by only $1\%$, they can significantly change some distributions in certain kinematic regions. For example, the relative NLO EW corrections to the muon transverse momentum distribution at 500 GeV amounts to $-20\%$. To illustrate how missing NLO EW corrections could masquerade as BSM physics, we show examples for the impact of dimension-8 operators in the SM Effective Field Theory framework on selected kinematic distributions.
\end{abstract}

\maketitle

\section{Introduction}
\label{sec:introduction}
The investigation of the self interactions of electroweak (EW) gauge bosons ($\gamma,Z,W^\pm$) at an
increasing level of precision offers a promising indirect window to Beyond-the-SM (BSM) physics, and is therefore an important goal of the CERN Large Hadron Collider (LHC). EW triboson production processes, $p\;p \to V\;V'\;V''$, where at least one gauge boson is a $W$ or $Z$ boson, are
especially interesting, since they provide direct access to the least tested EW gauge-boson interactions, the quartic-gauge-boson couplings (QGCs). For instance, $WWW$ production at the LHC, which directly probes the $WWWW$ QGC, was searched for by the ATLAS~\cite{ATLAS:2016jeu} and CMS~\cite{CMS:2019mpq} collaborations at $\sqrt{s}=8$ TeV and $13$ TeV, respectively, and only recently was observed for the first time by the ATLAS collaboration with a significance of $8.0\sigma$ using the full Run-2 data set~\cite{ATLAS:2022xnu}. The first evidence for a combination of heavy triboson production processes has been reported by the ATLAS collaboration in Ref.~\cite{ATLAS:2019dny} and the first observation of a combined $WWW, WWZ, WZZ$, and $ZZZ$ production signal was achieved by the CMS collaboration~\cite{CMS:2020hjs}.

While heavy triboson production processes have only recently become experimentally accessible, EW triboson processes
involving isolated photon(s) were among the first triboson cross section measurements performed at the LHC owing to their comparatively large cross sections ($W\gamma\gamma$ production has the largest inclusive cross section among the triboson processes with at least one heavy gauge boson). $W\gamma\gamma$ and $Z\gamma\gamma$ production cross sections have been measured by the CMS~\cite{CMS:2017tzy} and ATLAS~\cite{ATLAS:2015ify,ATLAS:2016qjc} collaborations at $\sqrt{s}=8$ TeV, and by the CMS collaboration~\cite{CMS:2021jji} at $\sqrt{s}=13$ TeV.
Evidence for $WW\gamma$ and $WZ\gamma$ productions at $\sqrt{s}=8$ TeV has been reported by both the CMS~\cite{CMS:2014cdf} and ATLAS~\cite{ATLAS:2017bon} collaborations. It is interesting to note that EW triboson production processes also allow for a study of triple-gauge-boson couplings (TGCs), complementing the ones performed in diboson production processes, and constitute an important background to direct BSM searches, especially in case of leptonic decays.

Clearly, to take full advantage of the potential of EW triboson processes at the LHC to search for indirect signals of BSM physics, SM predictions for the relevant observables need to be under superb theoretical control. In particular, the extraction of information about anomalous TGCs and QGCs or higher-dimensional operators in an Effective Field Theory (EFT) framework from measurements of kinematic distributions, requires the inclusion of both QCD and EW higher-order corrections. NLO QCD predictions for EW triboson processes with at least one $W$ or $Z$ boson and leptonic decays have been available for many years (see, e.g., Refs~\cite{Green:2016trm,Amoroso:2020lgh} for a review), and can readily be obtained for instance from the publicly available Monte Carlo (MC) program $\texttt{VBFNLO}$~\cite{Arnold:2008rz,Baglio:2014uba,Baglio:2011juf}. In recent years, the calculation of NLO EW corrections has also seen an increased activity which is not surprising given the importance of EW triboson production to the LHC physics program~\footnote{See also Ref.~\cite{Amoroso:2020lgh} for a discussion of the status of SM predictions and which calculations are still needed.}. Thanks to advances in the calculation of these corrections for processes with high-particle multiplicity in final states, the NLO EW predictions are gradually becoming more sophisticated, taken into account fully decayed final states without approximations. NLO EW corrections to $p\;p \to V\;V'\;V''$ processes with on-shell EW gauge bosons have been calculated in Refs.~\cite{Nhung:2013jta} ($WWZ$), \cite{Dittmaier:2017bnh} ($WWW$),
 \cite{Zhu:2020ous} ($WW\gamma$, including parton shower effects), and also served as benchmarks for $\texttt{MadGraph5\_aMC@NLOv3}$~\cite{Frederix:2018nkq} ($WWW$, $WZZ$, and $ZZZ$). Leptonic decays of the EW gauge bosons have been included in NLO EW predictions for $p\;p\to V\;V'\;V'' \to n \gamma +m$ leptons ($(n,m)=(0,6),(1,4),(2,2)$) in the narrow-width approximation for $WZZ$, $ZZZ$ and $ZZ\gamma$ production processes in Refs.~\cite{Shen:2015cwj, Wang:2016fvj, Wang:2017wsy} by adopting the $\texttt{MadSpin}$ method~\cite{Artoisenet:2012st}. NLO EW predictions for EW triboson production with leptonic decays, $p\;p\to n \gamma +m$ leptons, based on the complete set of Feynman diagrams for the $2\to (n+m)$-particle final state have been provided for $\gamma\gamma V$ ($V=W,Z$) production in Ref.~\cite{Greiner:2017mft} using $\texttt{Sherpa}$~\cite{Gleisberg:2008ta} and
$\texttt{GoSam}$~\cite{Cullen:2011ac, Cullen:2014yla}, and for $WWW$
 production in Refs.~\cite{Schonherr:2018jva,Dittmaier:2019twg} using $\texttt{RECOLA}$~\cite{Actis:2012qn,Actis:2016mpe} and $\texttt{OpenLoops}$~\cite{Buccioni:2019sur,Cascioli:2011va} \footnote{$\texttt{OpenLoops}$ also uses OPP reduction methods as implemented in $\texttt{CutTools}$ \cite{Ossola:2007ax} and $\texttt{OneLOop}$ \cite{vanHameren:2010cp} for the evaluation of one-loop scalar integrals.} as one-loop providers with $\texttt{COLLIER}$ \cite{Denner:2002ii,Denner:2005nn,Denner:2010tr,Denner:2016kdg} for the evaluation of one-loop scalar and tensor integrals.

The main focus of this paper is the calculation of NLO EW corrections to $p\;p\to e^{+}\;\nu_{e}\;\mu^{+}\;\mu^{-}\;\gamma$. This process includes the $W^{+}Z\gamma$ triboson process (with leptonic decays $W^+ \to e^+\;\nu_e$ and $Z\to \mu^+\;\mu^-$) and thus is sensitive to the $WWZ\gamma$ and $WW\gamma\gamma$ QGCs. The computation is based on the complete set of leading-order (LO) contributions of $\mathcal{O}(\alpha^5)$ in the electromagnetic coupling constant $\alpha$ and NLO EW contributions of $\mathcal{O}(\alpha^6)$. Considering the complete set of Feynman diagrams for this $\gamma+4l$ final state ensures that off-shell effects, spin correlations, and non-resonance contributions are fully taken into account. 

To our knowledge, the NLO EW corrections to $p\;p\rightarrow e^{+}\;\nu_{e}\;\mu^{+}\;\mu^{-}\;\gamma$ have not yet been studied in the literature. For completeness, and to study the numerical impact of different ways to combine NLO EW and NLO QCD corrections, we also calculated the ${\cal O}(\alpha_\text{s})$ corrections, where $\alpha_s$ denotes the strong coupling constant. We use the one-loop provider $\texttt{RECOLA}$ to calculate the LO contributions, virtual ${\cal O}(\alpha)$ and
${\cal O}(\alpha_s)$ corrections, while the real corrections are evaluated by $\texttt{MadDipole}$~\cite{Frederix:2008hu, Frederix:2010cj,Gehrmann:2010ry}, which is based on the dipole subtraction method~\cite{Catani:1996vz,Dittmaier:1999mb, Dittmaier:2008md}. Our emphasis is on studying the impact of these corrections on kinematic distributions and regions especially sensitive to effects of anomalous QGCs, applying experimentally inspired analysis cuts. These regions are often high-energy tails of distributions in the invariant mass or transverse momenta of final-state particles, and thus are known to potentially be considerably affected by NLO EW corrections due to the occurrence of EW Sudakov logarithms~\cite{Ciafaloni:1998xg}. To illustrate how missing NLO EW corrections could be mistaken as BSM effects, we choose as an example the SM Effective Field Theory (SMEFT) framework~\cite{Brivio:2017vri} to include dimension-8 operators and compare LO SMEFT with SM NLO EW predictions for a representative choice of distributions. 

The paper is organized as follows: In Section~\ref{sec:framework} we describe the calculational framework underlying our MC program, separately for the calculation of the virtual and real corrections in
Section~\ref{sec:virtual} and Section~\ref{sec:real} respectively. In Section~\ref{sec:validations} we provide a detailed description of the many checks we performed to validate the results of our MC program. In Section~\ref{sec:results}, after describing our choice of input parameters and of basic analysis cuts in Section~\ref{sec:inputs}, we provide numerical results for the total cross sections (Section~\ref{sec:tot}) and kinematic distributions (Section~\ref{sec:diff}) at NLO EW and NLO QCD together with a discussion of the residual theoretical uncertainty due to the factorization and renormalization scale variation at NLO QCD. The discussion of the numerical impact of NLO EW corrections closes with an example of the impact of dimension-8 operators in SMEFT in Section~\ref{sec:smeft}. Section~\ref{sec:summary} contains a brief summary and our conclusions.

\section{Calculational framework}
\label{sec:framework}

\begin{figure}[!tbp]
\centering
\begin{minipage}[b]{\textwidth}
    \includegraphics[width=\textwidth]{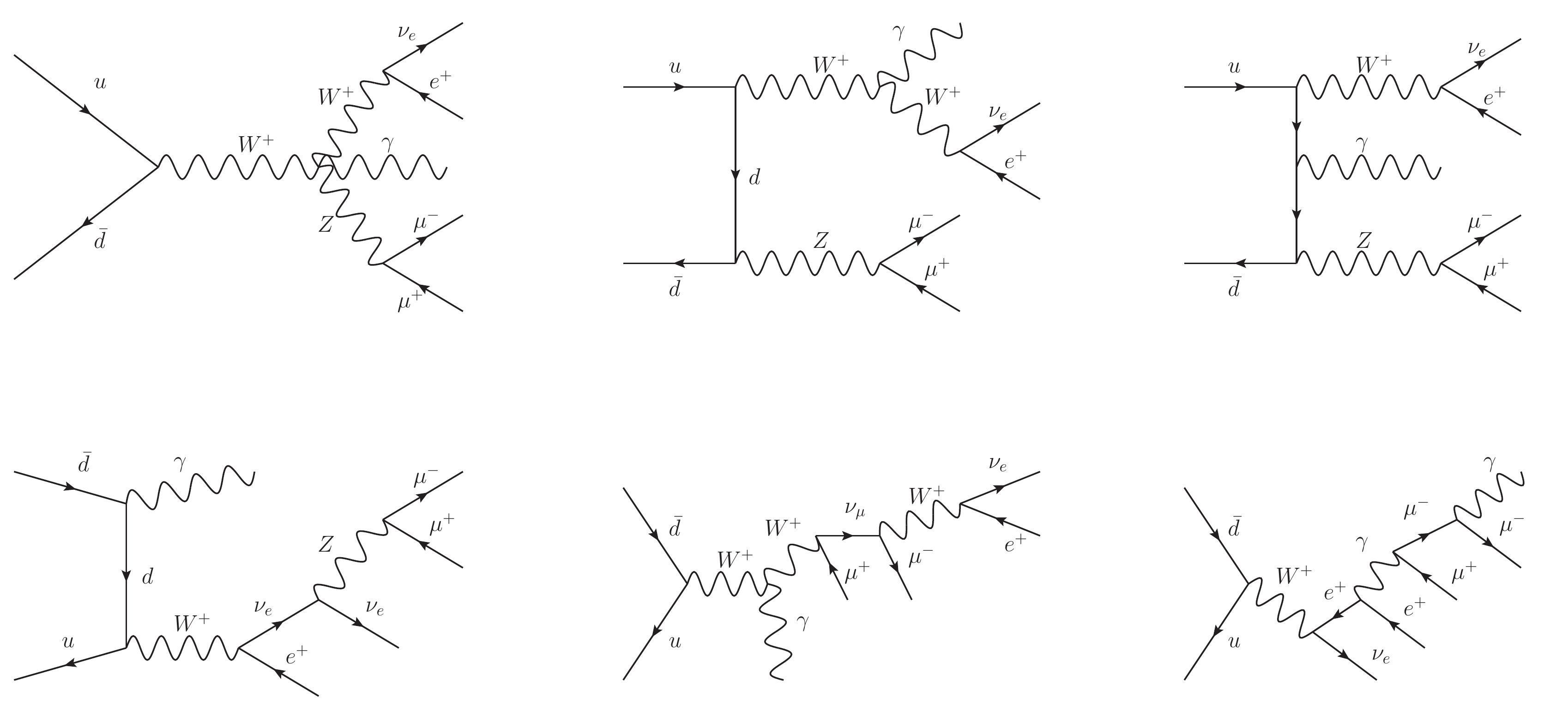}
\end{minipage}
\hfill
\caption{Sample LO Feynman diagrams for the $u\;\bar{d}\rightarrow e^{+}\;\nu_{e}\;\mu^{+}\;\mu^{-}\gamma$ process at ${\cal O}(\alpha^5)$.}
\label{fig:lo_feynman}
\end{figure}

The goal of this paper is to provide precise predictions for $W^{+}Z\gamma$ production with leptonic decays at the LHC with emphasis on calculating and studying the impact of EW ${\cal O}(\alpha)$ corrections to the process 
\begin{align}
    p\;p\rightarrow e^{+}\;\nu_{e}\;\mu^{+}\;\mu^{-}\;\gamma,
\end{align}
thereby taking into account the full off-shell effects, spin correlations and non-resonance contributions. We consider all fermions but the top quark to be massless.
For completeness, and to study the impact of different combinations of EW and QCD corrections, we also calculated ${\cal O}(\alpha_\text{s})$ corrections to this process. 
The calculation of the hadronic cross section is based on the master formula
\begin{align}
    \sigma(p_1,p_2)=\sum_{a,b}\int^{1}_{0} dx_{1} dx_{2} f_{a}(x_1,\mu_{\text{F}}) f_{b}(x_2,\mu_{\text{F}}) \hat{\sigma}_{ab} (p_a,p_b,\mu_{\text{F}},\mu_{\text{R}}),
\end{align}
where the sum is taken over all possible combinations of partons $a,b$ with momenta $p_{a,b}$  
determined by the fractions $x_{1,2}$ of the protons' momenta $p_{1,2}$. $f_{a,b}$ are the parton distribution functions (PDFs) depending on the momentum fractions $x_{1,2}$ and the factorization scale $\mu_{\text{F}}$. $\hat{\sigma}_{ab}$ denote the partonic cross sections which in case of NLO QCD also depend on the renormalization scale $\mu_{\text{R}}$. At NLO accuracy, $\hat{\sigma}_{ab}$ consists of Born contributions ($d\hat \sigma_{ab}^{\rm B}$), virtual one-loop corrections ($d\sigma^{\text{V}}_{ab}$) and real radiation corrections ($d\sigma^{\text{R}}_{ab}$):
\begin{align}
    \hat{\sigma}_{ab}^{\text{NLO}}=\int_{2\to 5}d\hat \sigma^{\text{B}}_{ab}+\int_{2\to 5}d\hat\sigma^{\text{V}}_{ab}+\int_{2\to 6}d\hat \sigma^{\text{R}}_{ab}.
\label{eq: master_formula_1}
\end{align}
As indicated, the Born and one-loop contributions are obtained by integrating over a $2\to 5$-particle phase space while the real corrections require an integration over a $2\to 6$-particle phase space. The partonic Born cross section is of $\mathcal{ O}(\alpha^5)$ and only receives contributions from the 
quark-induced processes:
\begin{align}
    q\;\bar{q}' \rightarrow e^{+}\;\nu_{e}\;\mu^{+}\;\mu^{-}\;\gamma,
\label{eq:partonic_process}
\end{align}
where $q$ and $q'$ denote the up-type light quark $(u,c)$ and the down-type light quark $(d,s)$ respectively. We do not include $b$-quark-initiated processes, since they have a negligible effect on the hadronic cross section due to the smallness of the $b$-quark PDF. In FIG.~\ref{fig:lo_feynman} we show a representative set of LO Feynman diagrams which consists of topologies arising from the $W^{+}Z\gamma$ production process with subsequent leptonic decays, some featuring QGCs and TGCs, and non-resonance contributions which do not arise from the $W^{+}Z\gamma$ production process. 
 
In our calculational framework, $d\hat\sigma_{ab}^{\text{B}}$ and $d\hat \sigma_{ab}^{\text{V}}$  are calculated by the one-loop provider $\texttt{RECOLA}$ and $d\hat \sigma_{ab}^{\text{R}}$ is evaluated by $\texttt{MadDipole}$. In the next two sections, we will discuss in more detail the calculations and validations of virtual one-loop corrections and real radiation corrections at both NLO EW and NLO QCD accuracy, and will also address some technical issues of the implementation of these tools in our MC framework.

\subsection{Virtual ${\cal O}(\alpha)$ and ${\cal O}(\alpha_\text{s})$ corrections}\label{sec:virtual}

\begin{figure}[!tbp]
\centering
\begin{minipage}[b]{\textwidth}
    \includegraphics[width=\textwidth]{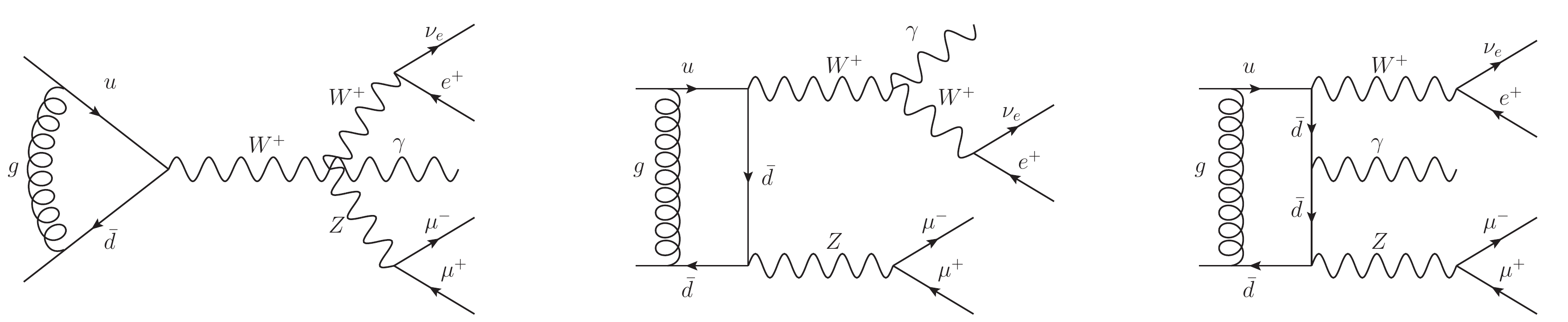}
\end{minipage}
\hfill
\caption{Sample Feynman diagrams for virtual ${\cal O}(\alpha_\text{s})$ corrections to $u\;\bar{d}\rightarrow e^{+}\;\nu_{e}\;\mu^{+}\;\mu^{-}\gamma$ at NLO QCD.}
\label{fig:qcd_virt_feynman}
\end{figure}

\begin{figure}[!tbp]
\centering
\begin{minipage}[b]{\textwidth}
    \includegraphics[width=\textwidth]{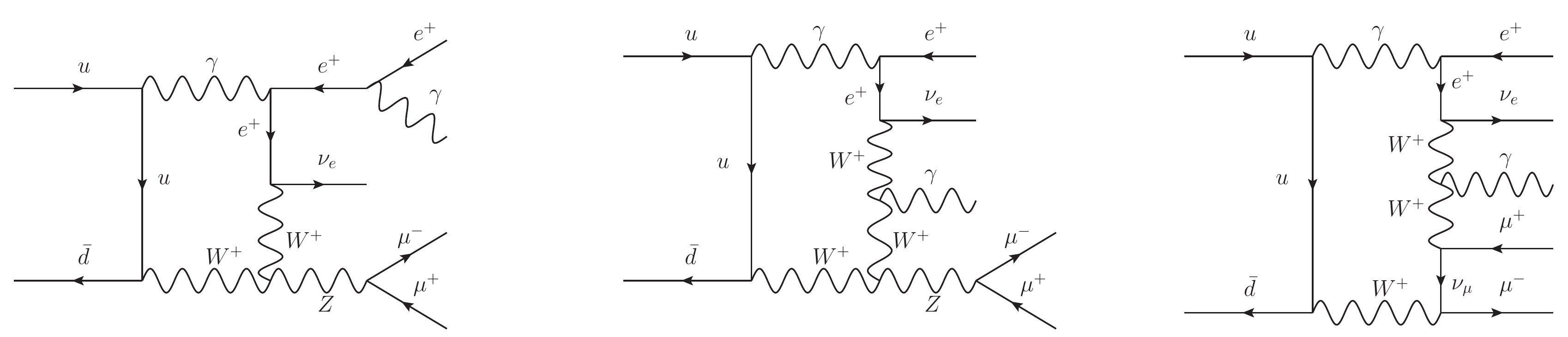}
\end{minipage}
\hfill
\caption{Sample Feynman diagrams for virtual ${\cal O}(\alpha)$ corrections to $u\;\bar{d}\rightarrow e^{+}\;\nu_{e}\;\mu^{+}\;\mu^{-}\gamma$ at NLO EW.}
\label{fig:ew_virt_feynman}
\end{figure}

The virtual ${\cal O}(\alpha_\text{s})$ corrections to the LO process $q\;\bar{q}' \rightarrow e^{+}\;\nu_{e}\;\mu^{+}\;\mu^{-}\;\gamma$ contain up to pentagon diagrams as shown in FIG.~\ref{fig:qcd_virt_feynman} where we provide some sample Feynman diagrams. The virtual ${\cal O}(\alpha)$ corrections to this process are more complicated, containing hexagon and heptagon diagrams which make their computation much more CPU costly than the virtual ${\cal O}(\alpha_\text{s})$ corrections~\footnote{For the process of Eq.~(\ref{eq:partonic_process}), the numerical evaluation of the interference of the one-loop amplitude and LO amplitude typically takes around 3 seconds for a single phase space point in case of ${\cal O}(\alpha)$ corrections and 10 microseconds in case of ${\cal O}(\alpha_{\text{s}})$ corrections on a 2.3 GHz Intel Core i5 processor.}. Sample Feynman diagrams for the  ${\cal O}(\alpha)$ corrections are displayed in FIG.~\ref{fig:ew_virt_feynman}. In our MC framework, the calculation of the interference of the one-loop amplitudes with the LO ones is performed by $\texttt{RECOLA}$ for both virtual ${\cal O}(\alpha)$ and ${\cal O}(\alpha_\text{s})$ corrections. 

Both ultraviolet (UV) and infrared (IR) divergences arising from the one-loop integrals are regularized using dimensional regularization in $d=4-2 \epsilon_{\text{UV,IR}}$ dimensions. The UV divergences are removed by renormalization in the complex-mass scheme~\cite{Denner:1999gp,Denner:2005fg,Denner:2006ic,Denner:2019vbn} in case of EW one-loop corrections and in the $\overline{\text{MS}}$-scheme for QCD one-loop corrections. In the complex-mass scheme the singularities appearing in the propagators of unstable particles ($W$ and $Z$ bosons and heavy quarks) when they tend to be on-shell are regulated in a gauge-invariant way. The aforementioned schemes are both implemented in $\texttt{RECOLA}$. The resulting renormalized one-loop contributions, $d\hat\sigma_{\text{V}}$ in Eq.~(\ref{eq: master_formula_1}) provided by $\texttt{RECOLA}$, still exhibit IR divergences which need to be cancelled by the counterparts in the real radiation corrections and the collinear PDF counterterms whenever IR-safe observables are calculated (see Section~\ref{sec:real}).

In the calculations of the NLO EW corrections, we chose as the EW input scheme the $G_\mu$-scheme \cite{Hollik:1988ii,Denner:1991kt,Andersen:2014efa}, where the electromagnetic coupling is determined from the Fermi constant $G_\mu$ and the pole masses of $W$ and $Z$ bosons as follows:
\begin{align}
       \alpha_{G_{\mu}}=\frac{\sqrt{2}}{\pi}G_{\mu}M_{\text{W}}^2\left( 1-\frac{M_{\text{W}}^2}{M_{\text{Z}}^2} \right).
\label{eq: alpha Gmu}
\end{align}
This choice has the advantage that large logarithmic corrections associated with the running of $\alpha$ are absorbed into the LO cross section. However, in processes with external photons at LO such as the one of Eq.~(\ref{eq:partonic_process}), and when considering light fermions to be massless throughout, the implementation of this scheme needs some care\footnote{The mixed scheme is discussed in detail in Ref.~\cite{Denner:2019vbn} and has been recently automated in  $\texttt{MadGraphG5\_aMC@NLOv3}$ for NLO EW corrections to processes with external photon(s) \cite{Pagani:2021iwa}. For completeness, we describe our implementation explicitly here.}. The UV counterterm contribution to the NLO EW corrections for the process of Eq.~(\ref{eq:partonic_process}) contains
the renormalization constant for the electric charge ($\delta Z_{e}$) and for the photon wave function ($\delta Z_{AA}$) as follows:
\begin{align}
\text{CT}_{\text{UV}} \propto \left[4 \delta Z_e+ \left( \delta Z_e+\frac{1}{2} \delta Z_{AA} \right) \right] \,  |\mathcal{M}_{\text{LO}}|^2,
\label{eq:uv_counterterm}
\end{align}
where the term $(\delta Z_e+\frac{1}{2} \delta Z_{AA})$ arises from the presence of an external photon and $\mathcal{M}_{\text{LO}}$ denotes the LO matrix element.  Both $\delta Z_e$ and $\delta Z_{AA}$ contain contributions from the derivative of the photon self energy, $\Pi^{AA}(0)$, evaluated at zero momentum, which exhibits large logarithmic contributions of the form $\log(m_f/\mu_{\text{R}})$ when retaining the light fermion masses $m_f$ in the calculation. In dimensional regularization, these logarithms manifest as single IR poles. In the $\alpha(0)$-scheme, $\delta Z_e$ is given by \cite{Hollik:1988ii,Denner:1991kt}
\begin{align}
    \delta Z_{e} \vert_{\alpha(0)}=\frac{1}{2}\Pi^{AA}(0)-\frac{s_{\text{W}}}{c_{\text{W}}}\frac{\Sigma^{AZ}_{\text{T}}(0)}{M_{\text{Z}}^2},
\end{align}
while in the $G_{\mu}$-scheme it takes the form
\begin{align}
    \delta Z_{e} \vert_{\alpha(G_{\mu})}=\frac{1}{2}\Pi^{AA}(0)-\frac{s_{\text{W}}}{c_{\text{W}}}\frac{\Sigma^{AZ}_{\text{T}}(0)}{M_{\text{Z}}^2}-\frac{1}{2}\Delta r,
\end{align}
where $\Delta r$ comprises the NLO EW corrections to muon decay, and reads~\cite{Sirlin:1980nh} 
\begin{align}
    \Delta r = \Pi^{AA}(0)-\frac{c^{2}_{\text{W}}}{s^{2}_{\text{W}}} \left( \frac{\Sigma^{ZZ}_{\text{T}}(M^{2}_{\text{Z}})}{M^{2}_{\text{Z}}}-\frac{\Sigma^{W}_{\text{T}}(M^{2}_{\text{W}})}{M^{2}_{\text{W}}} \right) + \frac{\Sigma^{\text{W}}_{\text{T}}(0)-\Sigma^{\text{W}}_{\text{T}}(M^{2}_{\text{W}})}{M^{2}_{\text{W}}} \nonumber\\
    +\frac{2c_{\text{W}}}{s_{\text{W}}}\frac{\Sigma^{AZ}_{\text{T}}(0)}{M^{2}_{\text{Z}}}+\frac{\alpha(0)}{4\pi s^{2}_{\text{W}}} \left( 6+\frac{7-4s^{2}_{\text{W}}}{2s^{2}_{\text{W}}}\ln c^{2}_{\text{W}} \right).
\label{eq: deltar}
\end{align}
As can be seen, in the $G_\mu$-scheme the $\Pi_{AA}$ contributions in $\delta Z_{e} \vert_{\alpha(G_{\mu})}$ cancel. However, if $\delta Z_{e} \vert_{\alpha(G_{\mu})}$ is used in Eq.~(\ref{eq:uv_counterterm}), the IR pole in $\delta Z_{AA}=-\Pi^{AA}(0)$ remains uncanceled. This can be avoided when using a mixed scheme where the electromagnetic coupling of the external photon at LO is taken as $\alpha(0)$, which alters the UV counterterm of Eq.~(\ref{eq:uv_counterterm}) into
\begin{align}
\text{CT}_{\text{UV}} \propto \left[ 4 \delta Z_e \vert_{\alpha(G_{\mu})}+ \left( \delta Z_e \vert_{\alpha(0)}+\frac{1}{2} \delta Z_{AA} \right) \right] |\mathcal{M}_{\text{LO}}|^2.
\end{align}
Since $\delta Z_{AA}=-\Pi^{AA}(0)$, the combination $(\delta Z_e \vert_{\alpha(0)}+\frac{1}{2} \delta Z_{AA})$
now remains IR finite. Evaluating $\alpha$ at zero momentum is also a more appropriate choice for an external photon coupling. To summarize, in this mixed scheme, the cross sections of our process are of order $\alpha^{4}_{G_{\mu}}\alpha(0)$ at LO, $\alpha^{4}_{G_{\mu}}\alpha(0)\alpha_\text{s}$ at NLO QCD and $\alpha^{5}_{G_{\mu}}\alpha(0)$ at NLO EW. To easily implement the mixed scheme in the calculation of the virtual EW corrections, we first calculated the renormalized one-loop contribution with $\texttt{RECOLA}$ in the pure $G_{\mu}$-scheme (of order $\alpha^6_{G_{\mu}}$) and then converted the results into the mixed scheme (of order $\alpha^{5}_{G_{\mu}}\alpha(0)$) as follows:
\begin{align}
    2 \text{Re} (\mathcal{M}_{\text{1-loop}}\mathcal{M}_{\text{LO}}^*) ( \alpha^{5}_{G_{\mu}}\alpha(0)) =  \frac{\alpha(0)}{\alpha_{G_\mu}} \cdot \left[2 \text{Re} (\mathcal{M}_{\text{1-loop}}\mathcal{M}_{\text{LO}}^*) ( \alpha^{6}_{G_{\mu}}) + \Delta r \cdot |\mathcal{M}_{\text{LO}}|^2 (\alpha^5_{G_\mu}) \right].
\end{align}

\subsection{Real ${\cal O}(\alpha)$ and ${\cal O}(\alpha_\text{s})$ corrections\label{sec:real}}

\begin{figure}[!tbp]
\centering
\begin{minipage}[b]{\textwidth}
    \includegraphics[width=\textwidth]{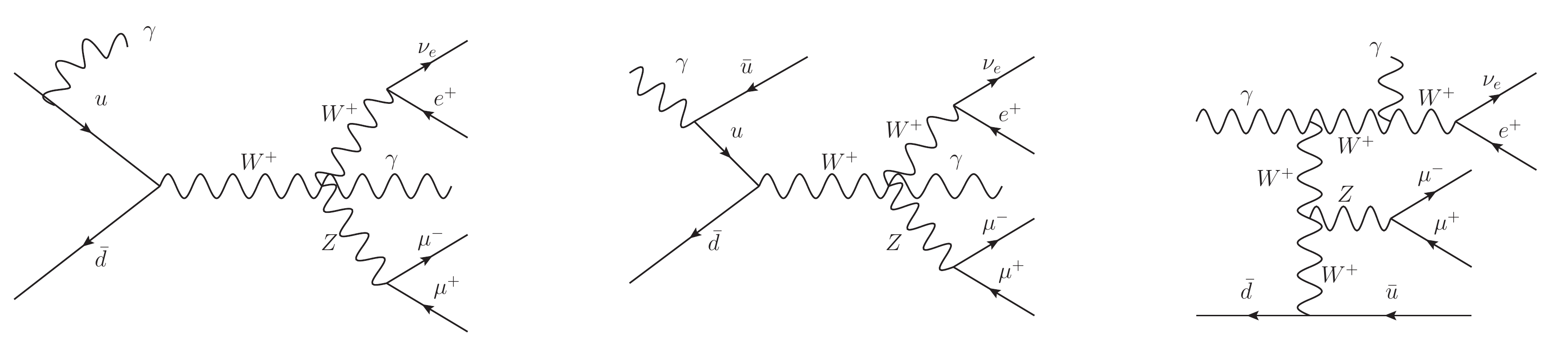}
\end{minipage}
\hfill
\caption{Sample Feynman diagrams of real $\mathcal{O}(\alpha)$ corrections and photon-induced corrections to $u\;\bar{d}\rightarrow e^{+}\;\nu_{e}\;\mu^{+}\;\mu^{-}\gamma$ at NLO EW.}
\label{fig:ew_real_feynman}
\end{figure}

\begin{figure}[!tbp]
\centering
\begin{minipage}[b]{\textwidth}
    \includegraphics[width=\textwidth]{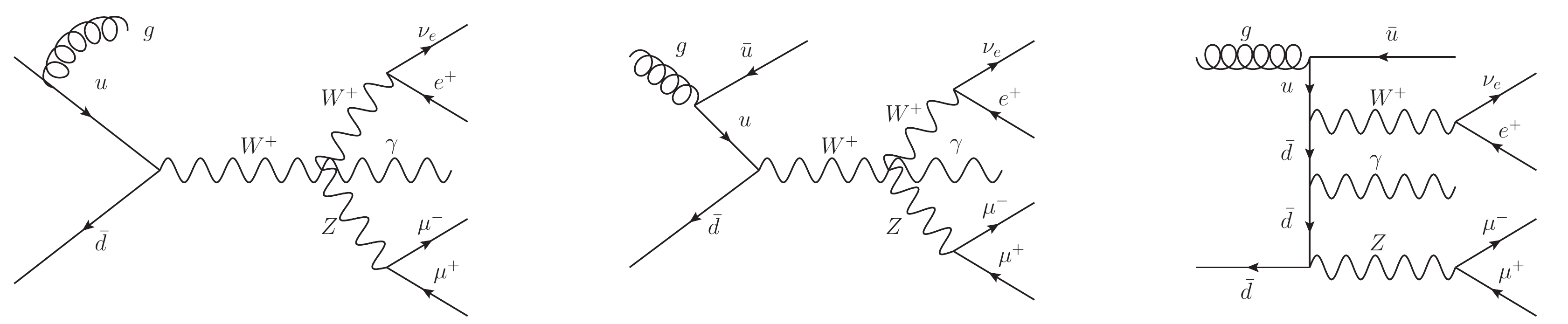}
\end{minipage}
\hfill
\caption{Sample Feynman diagrams of real $\mathcal{O}(\alpha_\text{s}$) corrections and gluon-induced corrections to $u\;\bar{d}\rightarrow e^{+}\;\nu_{e}\;\mu^{+}\;\mu^{-}\gamma$ at NLO QCD.}
\label{fig:qcd_real_feynman}
\end{figure}

The EW real corrections at ${\cal O}(\alpha)$ contain the quark-induced processes, where a photon can be emitted from either initial-state quarks, final-state charged leptons or a $W$ boson, and the photon-induced processes. Similarly, the QCD real corrections at ${\cal O}(\alpha_\text{s})$ are comprised of gluon radiation off initial-state quarks in the quark-induced processes and the gluon-induced processes, featuring initial-state gluon splitting into a $q\bar q$ pair. Sample Feynman diagrams of EW and QCD real corrections are shown in FIG.~\ref{fig:ew_real_feynman} and FIG.~\ref{fig:qcd_real_feynman}, respectively. The arising IR divergences of soft and collinear origin 
have to be cancelled against those in the virtual corrections in IR-safe observables. The IR singularities are extracted by using the dipole subtraction method implemented in $\texttt{MadDipole}$ \footnote{We noticed that in the public version of $\texttt{MadDipole}\;\text{(v-4.5.1)}$ the finite part of the QED integrated dipoles needed a correction. This has been confirmed by one of the authors who provided us with a private, corrected version of the code.}, and removed by combining them with the virtual corrections and the collinear PDF counterterm according to the following master formula for the quark-induced partonic cross section:
\begin{align}
    \hat{\sigma}_{q\bar q'}=\int_{2\to 5}d\sigma^{\text{B}}_{q\bar q'}+\int_{2\to 5} \left( d\sigma^{\text{V}}_{q\bar q'} + d\sigma^{\text{I}}_{q\bar q'} \right)+\int_{2\to 6} \left( d\sigma^{\text{R}}_{q\bar q'}-d\sigma^{\text{A}}_{q\bar q'} \right) +\int_{2 \to 5}d\sigma^{\text{C}}_{q\bar q'}.
\end{align}
The differential dipoles $d\sigma^{\text{A}}_{q\bar q'}$ match the singular behaviour of real corrections $d\sigma^{\text{R}}_{q\bar q'}$ locally, and the IR poles in the virtual corrections are cancelled upon combining with the integrated dipoles $d\sigma^{\text{I}}_{q\bar q'}$. The collinear PDF counterterms $d\sigma^{\text{C}}_{q\bar q'}$ absorb the residual initial-state collinear singularities into PDFs in the $\overline{\text{MS}}$-factorization scheme. Both QCD and QED dipole contributions, $d\hat \sigma^{\text{A,I}}_{q\bar q'}$, along with the collinear PDF counterterms are generated by $\texttt{MadDipole}$. In order for an exact pole cancellation to happen and a proper combination of the finite contributions, the conventions for the prefactors of the Laurent expansion about the IR poles in $d$ dimensions in the integrated dipoles and virtual one-loop corrections have to coincide. The tools we rely on to calculate the virtual one-loop corrections and integrated dipoles do have different conventions, and thus an additional adjustment is required. The convention used by $\texttt{RECOLA}$ for the expansion is~\cite{Actis:2016mpe}
\begin{align}
    (4\pi^{\epsilon_\text{IR}})\Gamma(1+\epsilon_\text{IR}) \left(\frac{A}{\epsilon_\text{IR}^2}+\frac{B}{\epsilon_\text{IR}}+C \right),
\end{align}
and the convention used by $\texttt{MadDipole}$ is~\cite{Frederix:2010cj}
\begin{align}
    \frac{(4\pi)^{\epsilon_\text{IR}}}{e^{\gamma\epsilon_\text{IR}}}\left(\frac{A}{\epsilon_\text{IR}^2}+\frac{B}{\epsilon_\text{IR}}+C' \right),
\end{align}
where $A$ and $B$ denote the double and single pole coefficients, respectively, while $C$ and $C'$ denote the finite contributions, and $\gamma$ is the Euler-Mascheroni constant with $\gamma\approx0.5722$. The difference between the conventions does not affect double and single pole coefficients but finite contributions. In our calculation, we adopt the $\texttt{RECOLA}$ convention and convert the finite contribution of the integrated dipoles computed by $\texttt{MadDipole}$ accordingly as follows
\begin{align}
    C'\rightarrow C'-\frac{\pi^{2}}{12}A^{2}.
\end{align}
The gluon/photon-induced processes, which contribute for the first time at NLO, only exhibit a collinear singularity due to initial-state gluon/photon splitting into a $q\bar q$ pair. Therefore, the master formula for these processes simplifies to  
\begin{align}
    \hat{\sigma}_{g/\gamma\;q}=\int _{2\to 5} d\sigma^{\text{I}}_{g/\gamma\;q} +\int_{2\to 6} \left( d\sigma^{\text{R}}_{g/\gamma\;q}-d\sigma^{\text{A}}_{g/\gamma\;q} \right) +\int_{2\to 5}d\sigma^{\text{C}}_{g/\gamma\;q}.
\end{align}
and only collinear subtraction terms and PDF counterterms are needed.

\subsection{Validations \label{sec:validations}}

To validate the NLO EW and NLO QCD calculations of our MC framework, we performed numerous checks. Unless noted otherwise, we used the setup described in Section~\ref{sec:inputs}. We compared the LO and real radiation squared amplitudes with those calculated by $\texttt{MadGraph5}$ \cite{Alwall:2014hca} at several phase space points. The finite contributions, coefficients of double and single poles of QCD and EW one-loop corrections have been compared with those calculated by $\texttt{MadLoop3}$ \cite{Frederix:2018nkq} and $\texttt{OpenLoops2}$ \cite{Buccioni:2019sur}, again at several phase space points. We found good agreement in all comparisons. To verify the proper implementation of the dipole subtraction method using \texttt{MadDipole}, we checked the cancellations of the double and single IR poles among the virtual contributions, the integrated dipoles and the collinear PDF counterterms. We also checked that the real squared amplitudes in the soft/collinear limits indeed approach the value of the differential dipoles. As an example, we show in Appendix~\ref{app:pole} (Table \ref{table: EW pole cancellations}) the cancellations of the IR poles in the case of NLO EW corrections at a single phase-space point (provided in Table \ref{table: phase space point}).

\begin{figure}[!tbp]
\centering
\begin{minipage}[b]{0.49\textwidth}
    \includegraphics[width=\textwidth]{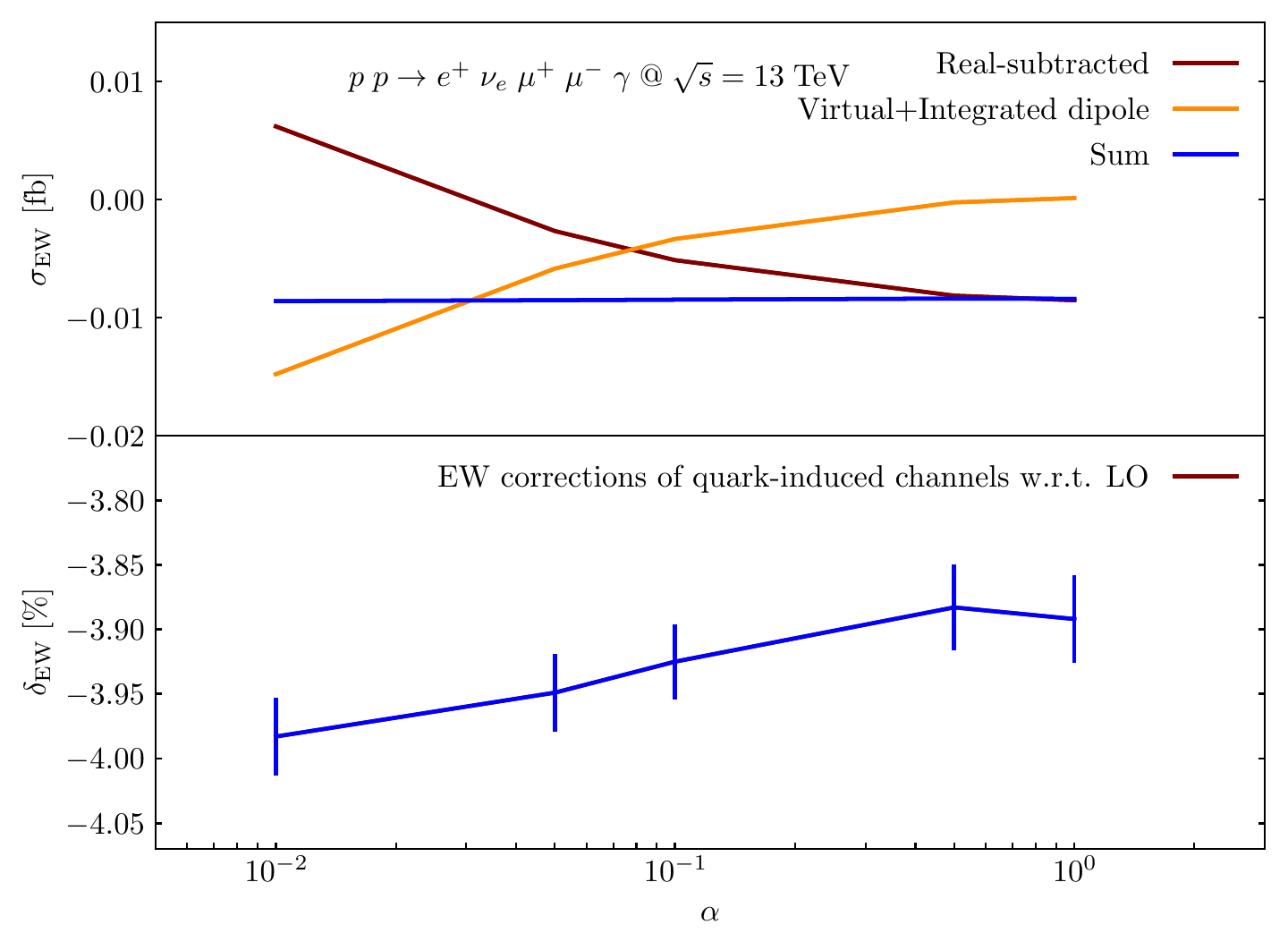}
\end{minipage}
\hfill
\begin{minipage}[b]{0.49\textwidth}
    \includegraphics[width=\textwidth]{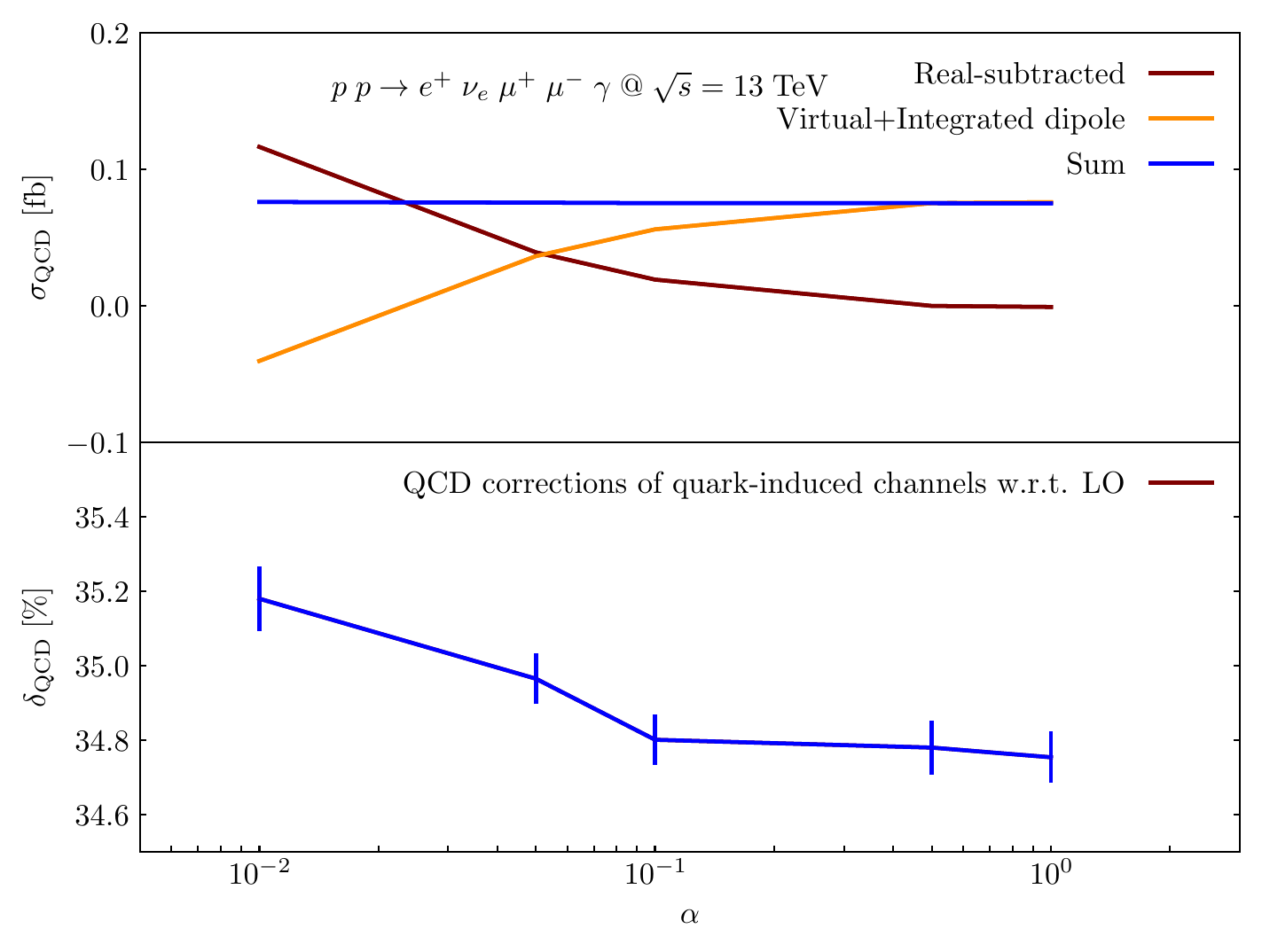}
\end{minipage}
\caption{The $\alpha$-parameter dependence of total hadronic cross sections at NLO EW (left) and NLO QCD (right) for the process $u\;\bar{d}\rightarrow e^{+}\;\nu_{e}\;\mu^{+}\;\mu^{-}\;\gamma$, evaluated with the QED and QCD dipole subtraction method implemented in MadDipole. The top panels show the real-subtracted corrections (red), the virtual+integrated dipoles (orange) and their sum (blue) with $\alpha=0.01, 0.05, 0.1, 0.5$ and 1. The bottom panels show the relative corrections with respect to the LO result.}
\label{fig:alpha_dependence}
\end{figure}

At the total hadronic cross section level, the $\alpha$-parameter dependence \cite{Nagy:2003tz} is checked. This parameter is introduced to restrict the phase space for real radiation where dipole subtraction is needed: $\alpha=1$ corresponds to no restriction, i.e. all  dipoles are subtracted in the entire phase space after the application of kinematic cuts. A smaller $\alpha$-parameter means that only dipoles are subtracted that mimic the singular behavior of the real corrections in this phase space region. As a result,
a finite contribution is shifted between real-subtracted corrections and integrated dipoles, but their sum has to be $\alpha$-independent. We checked the $\alpha$-parameter dependence for NLO QCD and NLO EW corrections to the quark-induced process as shown in FIG.~\ref{fig:alpha_dependence}. In the top panels, we show the hadronic total cross sections including the real-subtracted corrections (red) and the virtual-plus-integrated-dipoles corrections (orange), as well as their sum (blue). In the bottom panels, we show the relative corrections of the sum with respect to the LO result. 
The $\alpha$-parameter independence can be seen with $\alpha=1,\;0.5,\;0.1,\;0.05$, while the case of $\alpha=0.01$ does introduce a small dependence, i.e. about $0.05\%$ for NLO EW corrections and about $0.2\%$ for NLO QCD corrections. However, this effect will have no noticeable impact given the large QCD scale uncertainty which will be discussed in detail in Section~\ref{sec:results}. 

\begin{figure}[!tbp]
\centering
\begin{minipage}[b]{0.49\textwidth}
    \includegraphics[width=\textwidth]{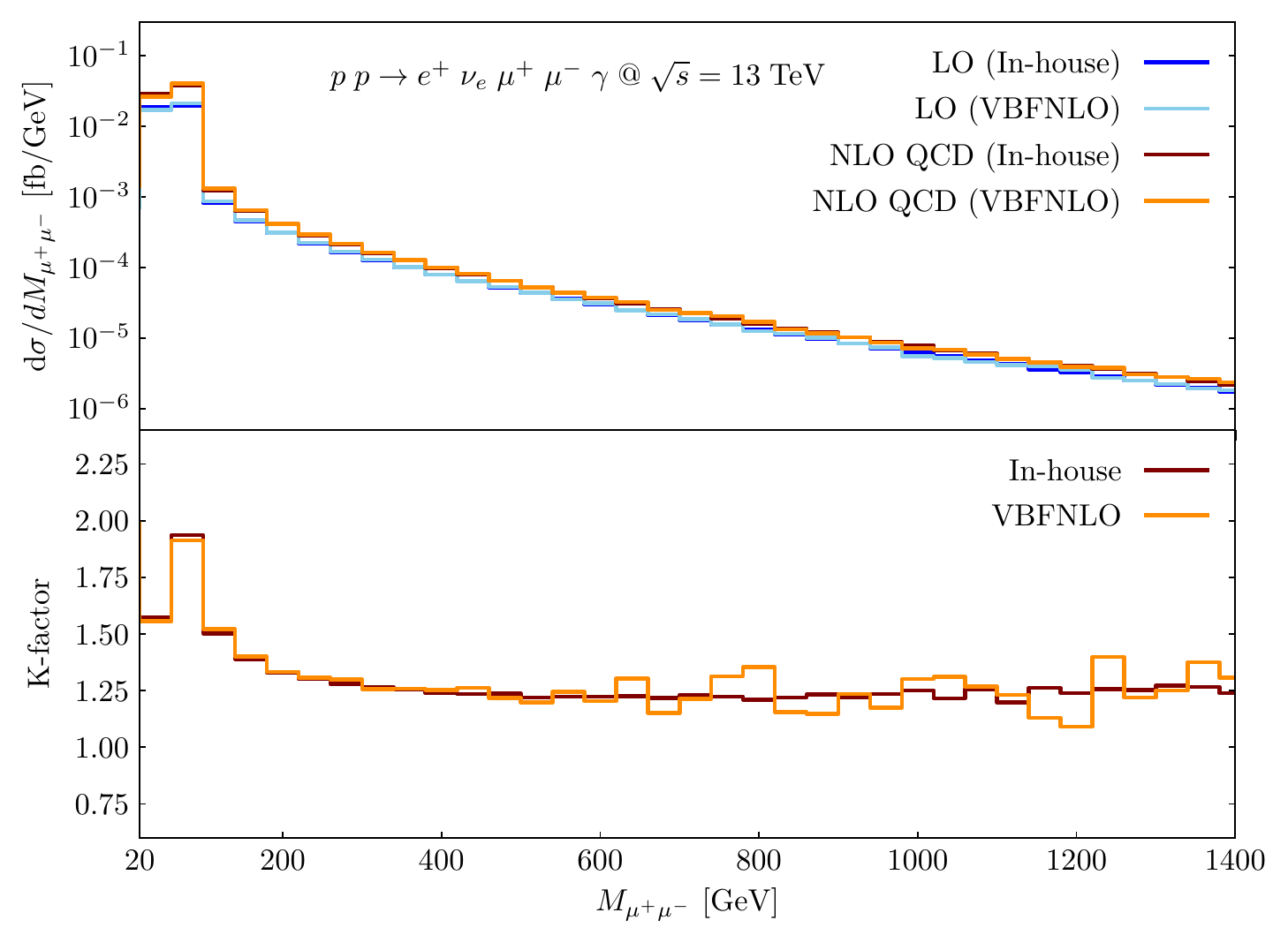}
\end{minipage}
\hfill
\begin{minipage}[b]{0.49\textwidth}
    \includegraphics[width=\textwidth]{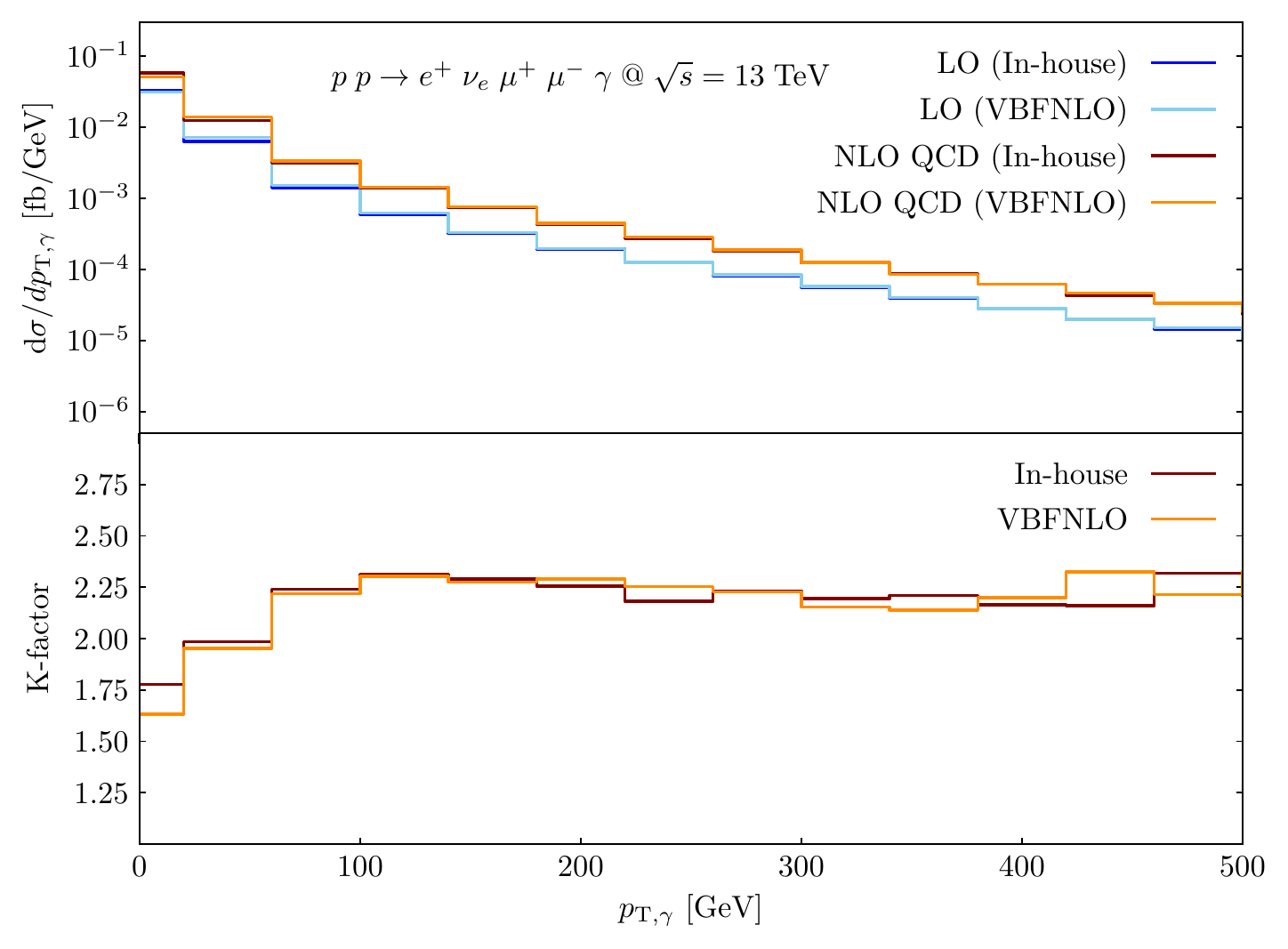}
\end{minipage}
\caption{The LO results and NLO QCD corrections to the distribution of the invariant mass of the $\mu^+\mu^-$ pair (left) and the transverse momentum of the isolated photon (right), calculated by our MC program and $\texttt{VBFNLO-2.7.0}$. The corresponding K-factors are shown in the bottom panel of each plot.}
\label{fig:valid_vbfnlo}
\end{figure}

We also performed a comparison of kinematic distributions at NLO QCD with those computed by $\texttt{VBFNLO}$ \cite{Arnold:2008rz}. We produced results for the invariant mass of the $\mu^+\mu^-$ pair and the transverse momentum of the isolated photon. As can be seen in FIG.~\ref{fig:valid_vbfnlo}, there is good agreement within the statistical MC uncertainties of the two MC programs.

Finally, we recalculated some results for total hadronic cross sections available in the literature (adjusting our input parameters and cuts accordingly): NLO QCD corrections to on-shell $WWZ$ production \cite{Nhung:2013jta}, NLO EW corrections to the neutral-current Drell-Yan process ($\delta^{\text{rec}}_{q\bar{q},\text{phot}}$ and $\delta_{q\bar{q},\text{weak}}$ for $M_{ll}>50$ GeV of Table~1 in \cite{Dittmaier:2009cr}), and NLO EW corrections to $Z\gamma$ production with leptonic decays ($\delta_{\text{phot}}^{\text{CS}}$, $\delta_{\text{weak},q\bar{q}}$ and $\delta_{\gamma\gamma}$ of Table~1 in \cite{Denner:2015fca}). The comparison with the results obtained with our MC program is shown in Appendix~\ref{app:recalculated} (Tables~\ref{table:wwz fixed scale},\ref{table:wwz dynamic scale},\ref{table:e+e-} and \ref{table:e+e-a}). In general we found good agreement within the statistical uncertainties of the MC programs. Small differences are at most at the $0.9\%$ level of the relative correction which is not surprising when comparing different MC implementations of higher-order corrections.

\section{Numerical results}
\label{sec:results}

In this section we present results for the total cross sections and kinematic distributions for $p\;p\to e^{+}\;\nu_{e}\;\mu^{+}\;\mu^{-}\;\gamma$ at the 13 TeV LHC for a basic set of analysis cuts, and discuss the impact of NLO EW and NLO QCD together with an assessment of the residual renormalization and factorization scale uncertainty. We also study the impact of different ways to combine NLO EW and NLO QCD corrections and compare the effect of NLO EW corrections with those of dimension-8 operators in SMEFT.

\subsection{Input parameters and analysis cuts \label{sec:inputs}}

In the numerical evaluation we use the following on-shell masses and widths~\cite{ParticleDataGroup:2020ssz}:
\begin{align}
        m_{\rm{W}}&=80.379\;\rm{GeV}, & \Gamma_{\rm{W}}&=2.085\;\rm{GeV}, \nonumber\\
        m_{\rm{Z}}&=91.1876\;\rm{GeV}, & \Gamma_{\rm{Z}}&=2.4952\;\rm{GeV}, \nonumber\\
        m_{\rm{H}}&=125.0\;\rm{GeV}, & m_{\rm{t}}&=173.1\;\rm{GeV}.
\label{eq: on-shell masses}
\end{align}
All fermions but the top quark are considered massless. Since the top quark and Higgs boson widths are not needed we have set them to zero. Since in \texttt{RECOLA} the complex-mass scheme is employed, the pole masses and widths are used throughout the calculations which are converted from the on-shell ones via~\cite{Bardin:1988xt}
\begin{align}
        M_{\text{Pole}}&=\frac{M_{\text{OS}}}{\sqrt{1+(\Gamma_{\text{OS}}/M_{\text{OS}})^2}}, \label{eq:pole mass}\\
        \Gamma_{\text{Pole}}&=\frac{\Gamma_{\text{OS}}}{\sqrt{1+(\Gamma_{\text{OS}}/M_{\text{OS}})^2}}.
\label{eq:pole width}
\end{align}
At LO and NLO EW, our considered process is of order $\alpha^{4}_{G_{\mu}}\alpha(0)$ and $\alpha^{5}_{G_{\mu}}\alpha(0)$ respectively, where $\alpha(0)$ is associated with the external hard photon and is taken to be $\alpha(0)=1/137.035999084$ \cite{ParticleDataGroup:2020ssz}.
All other EW couplings including those related to the initial-state photon splittings and final-state photon radiations are determined in the $G_{\mu}$-scheme as discussed in Section~\ref{sec:virtual}. With $G_{\mu}=1.1663787\times 10^{-5}\;\text{GeV}^{-2}$~\cite{ParticleDataGroup:2020ssz} and the pole masses of Eq.~(\ref{eq:pole mass}) one finds $\alpha_{G_{\mu}}=1/132.30808053$ using Eq.(\ref{eq: alpha Gmu}). Through an interface with the \texttt{LHAPDF6} library \cite{Buckley:2014ana}, we employ the \texttt{NNPDF31\_nlo\_as\_0118\_luxqed} set \cite{Bertone:2017bme} for both LO and NLO calculations in which the photon PDF and QED effects, for example in the DGLAP evolution, are included. The strong coupling constant $\alpha_\text{s}$ only enters at NLO QCD and is determined in accordance with the chosen PDF set ($\alpha_\text{s}(m_{\text{Z}})=0.118$ and its running value is extracted from the PDF set considering five light flavors).
For the renormalization ($\mu_\text{R}$)  and factorization ($\mu_\text{F}$) scales, their central values $\mu_0$ are set equal and defined as the invariant mass of the final-state particles:
\begin{align}
    \mu_{0}=\mu_{\text{R}}=\mu_{\text{F}}\equiv \sqrt{(p_{e^{+}}+p_{\nu_{e}}+p_{\mu^{+}}+p_{\mu^{-}}+p_{\gamma})^2}.
\label{eq:scale_def}
\end{align}
To assess the scale uncertainty of the NLO QCD cross sections we performed a seven-point scale variation. The seven scale choices for $(\mu_\text{R},\mu_\text{F})$ are $(\mu_0,\mu_0)$, $(2\mu_0,2\mu_0)$, $(0.5\mu_0,0.5\mu_0)$, $(2\mu_0,\mu_0)$, $(0.5\mu_0,\mu_0)$, $(\mu_0,2\mu_0)$ and $(\mu_0,0.5\mu_0)$. 

In order to obtain well-defined cross sections, we performed a photon-charged-lepton recombination procedure and apply a basic set of analysis cuts, loosely inspired by experimental analysis cuts for triboson production processes at the LHC.  Photon-charged-lepton recombination is needed in the calculation of the EW real corrections with two photons in the final state where one photon can be collinear to a final-state charged lepton. The recombination procedure is applied so that these regions of phase space are treated fully inclusively even in the presence of lepton identification cuts.  All other contributions only contain one photon in the final state which will be identified by applying a photon isolation cut and thus can never become collinear to a final-state charged lepton. When applying the recombination procedure to the real EW corrections, e.g., to $u\;\;\bar{d}\rightarrow e^{+}\;\;\nu_{e}\;\;\mu^{+}\;\;\mu^{-}\;\;\gamma\;\;\gamma$, the separations of the photons and the charged leptons in the pseudorapidity-azimuthal-angle plane, 
\begin{align}
R_{ij}=\sqrt{\Delta \eta_{ij}^{2}+\Delta\phi_{ij}^{2}}
\label{eq: angular separation}
\end{align}
are calculated, where $j=\gamma$ and $i=l\in \{e^{+},\mu^{+},\mu^{-}\}$, $\Delta \eta_{i\gamma}=\eta_i-\eta_\gamma$
is the pseudorapidity difference and $\Delta\phi_{i\gamma}=\phi_i-\phi_\gamma$ is the corresponding azimuthal angle difference. The photon and charged lepton with the smallest $R_{l\gamma}$ are recombined, i.e. their four-momenta are added, as long as $R_{l\gamma}<0.1$. In case of $R_{l\gamma}<0.1$ for both photons, the event will be rejected. If no recombination takes place, the harder photon satisfying $p_{\text{T},\gamma}>15$ GeV and $|\eta_{\gamma}|<2.5$ will be labeled as the identified photon. 

In the gluon-induced and photon-induced processes, as well as the QCD real corrections to the quark-induced processes, the final-state parton and photon may become collinear and thus induce extra IR-singularities of QED origin. To exclude this region of phase space two methods are commonly employed: democratic clustering with the help of a quark-to-photon fragmentation function \cite{Glover:1993xc} or Frixione isolation \cite{Frixione:1998jh}. In this paper, we choose using Frixione isolation to avoid having to introduce a fragmentation function dependence in our predictions. A comparison of the impact of EW and QCD corrections in $l^+l^-\gamma$ production in Ref. \cite{Denner:2015fca} when using either method has shown no difference in case of EW corrections and QCD corrections only differ by about $0.5\sim1\%$. Our results are based on the Frixione isolation cut applied as follows: The event is accepted if $R_{i\gamma}>\delta_{0}$ (Eq.~(\ref{eq: angular separation}) with $i=q,g$). In the case of $R_{i\gamma}<\delta_{0}$, the event is accepted only if 
\begin{align}
    p_{\text{T},i}\le \varepsilon\; p_{\text{T},{\gamma}}\frac{1-\cos R_{i\gamma}}{1-\cos\delta_{0}},
\end{align}
where $\delta_{0}$ the isolation cone size and $\varepsilon$ the damping parameter. We chose $\delta_{0}=0.7$ and $\varepsilon=1$.

After the application of the photon recombination procedure and the Frixione isolation cut, we applied the following additional analysis cuts: For the transverse momentum and pseudorapidity of the final-state photon and charged leptons, as well as the missing transverse momentum, we require
\begin{align}
p_{\text{T},\gamma}&>15\;\rm{GeV}, & |\eta_{\gamma}|&<2.5, & p_{\text{T},l}&>20\;\rm{GeV}, & |\eta_{l}|&<2.5, & \slashed{p}_{\text{T}}>15\;\text{GeV}.
\end{align}
The angular separations between photon and charged leptons along with the invariant mass of the $\mu^+\mu^-$ pair have to satisfy
\begin{align}
    R_{l\gamma}&>0.4, & m_{\mu^{+}\mu^{-}}>20\;\text{GeV},
\end{align}
which ensures that there is no collinear singularities from final-state photon splitting into leptons or photons radiating off the charged leptons at LO. Here, we do not impose restrictions on the angular separations between charged leptons ($R_{ll}$) or perform recombinations for two photons (when $R_{\gamma\gamma}<0.1$) given that they are not technically required.

\subsection{Total cross sections at NLO EW and NLO QCD}
\label{sec:tot}

We present the results for the total cross sections at LO, NLO QCD and NLO EW for the process $p\;p\rightarrow e^{+}\;\nu_{e}\;\mu^{+}\;\mu^{-}\;\gamma$ at $\sqrt{s}=13$ TeV in Table~\ref{table:total cross sections}. We also provide the relative NLO EW correction, $\delta_{\text{EW}}=\sigma_{\text{EW}}/\sigma_{\text{LO}}-1$, and the QCD $K$-factor, $K=\sigma_{\text{QCD}}/\sigma_{\text{LO}}$.
In case of NLO EW corrections, we show the quark-induced and photon-induced contributions separately, where the quark-induced contribution ($\delta_{\text{EW}}^{q\bar{q}}$) and the photon-induced contribution ($\delta_{\text{EW}}^{\gamma q(\bar{q})}$) have opposite signs and cause a large cancellation. The NLO EW corrections are therefore negligible compared to the large NLO QCD corrections at the total cross section level. However, in the next section, we will see that NLO EW corrections can have a significant impact on various kinematic distributions.
\begin{table*}[!ht]
\begin{tabular}{|c|c|c|c|c|c|c|}
\hline
        $\sigma_{\text{LO}}$ [fb] & $\sigma_{\text{QCD}}$ [fb] & $K$-factor & $\sigma_{\text{EW}}$ [fb] & $\delta_{\text{EW}}$ [$\%$] & $\delta_{\text{EW}}^{q\bar{q}}$ [$\%$] & $\delta_{\text{EW}}^{\gamma q(\bar{q})}$ [$\%$] \\
\hline
\hline
        $0.20869(5)$ & $0.3588^{+3.90\%}_{-3.23\%}(2)$ & 1.719(1) & 0.2101(1) & 0.97(1) & -3.99(4) & +4.96(1) \\
\hline
\end{tabular}
\caption{Total cross sections and relative corrections for $p\;p\rightarrow e^{+}\;\nu_{e}\;\mu^{+}\;\mu^{-}\;\gamma$ at LO ($\sigma_{\text{LO}}$), NLO QCD ($\sigma_{\text{QCD}}$) and NLO EW ($\sigma_{\text{EW}}$) at $\sqrt{s}=13$ TeV. The statistical uncertainties of the MC integration are reported in the last digits. The scale uncertainties are indicated as the upper and lower limits of the NLO QCD cross section. Separate results are also shown for the quark-induced and photon-induced contributions to the relative NLO EW corrections, $\delta_{\text{EW}}^{q\bar{q}}$ and $\delta_{\text{EW}}^{\gamma q(\bar{q})}$ respectively.}
\label{table:total cross sections}
\end{table*}
\begin{table*}[!ht]
\begin{tabular}{|c|c|c|c|c|c|c|}
\hline
        $(m,n)$ & (2,2) & (0.5,0.5) & (2,1) & (0.5,1) & (1,2) & (1,0.5) \\
\hline
\hline
        $\sigma_{\text{QCD}}$ [fb] & 0.3472(2) & 0.3728(2) & 0.3472(2) & 0.3716(3) & 0.3570(3) & 0.3577(3)  \\
        $\delta [\%]$ & -3.233 & +3.902 & -3.233 & +3.567 & -0.502 & -0.307 \\
\hline
\end{tabular}
\caption{Total NLO QCD cross sections for $p\;p\rightarrow e^{+}\;\nu_{e}\;\mu^{+}\;\mu^{-}\;\gamma$ at $\sqrt{s}=13$ TeV for seven scale choices for the renormalization and factorization scale, $(\mu_{\text{R}},\mu_{\text{F}})=(m,n) \cdot \mu_0$, where $(m,n)$ are the multiples of the central value $\mu_0$ of Eq.~(\ref{eq:scale_def}). We also provide the relative difference to the NLO QCD total cross section at the central scale $\sigma_{\text{QCD}}^{(1,1)}=0.3588$~fb, i.e. $\delta=\sigma_{\text{QCD}}^{(m,n)}/\sigma_{\text{QCD}}^{(1,1)}-1$.}
\label{table: scale variations}
\end{table*}
In Table~\ref{table: scale variations} we show the results of a 
seven-point scale variation to assess the scale uncertainty of the NLO QCD total cross sections as shown in Table~\ref{table:total cross sections}.

\subsection{Differential cross sections at NLO EW and NLO QCD}\label{sec:diff}

\begin{figure}[!tbp]
\centering
    \begin{minipage}[b]{0.49\textwidth}
    \includegraphics[width=\textwidth]{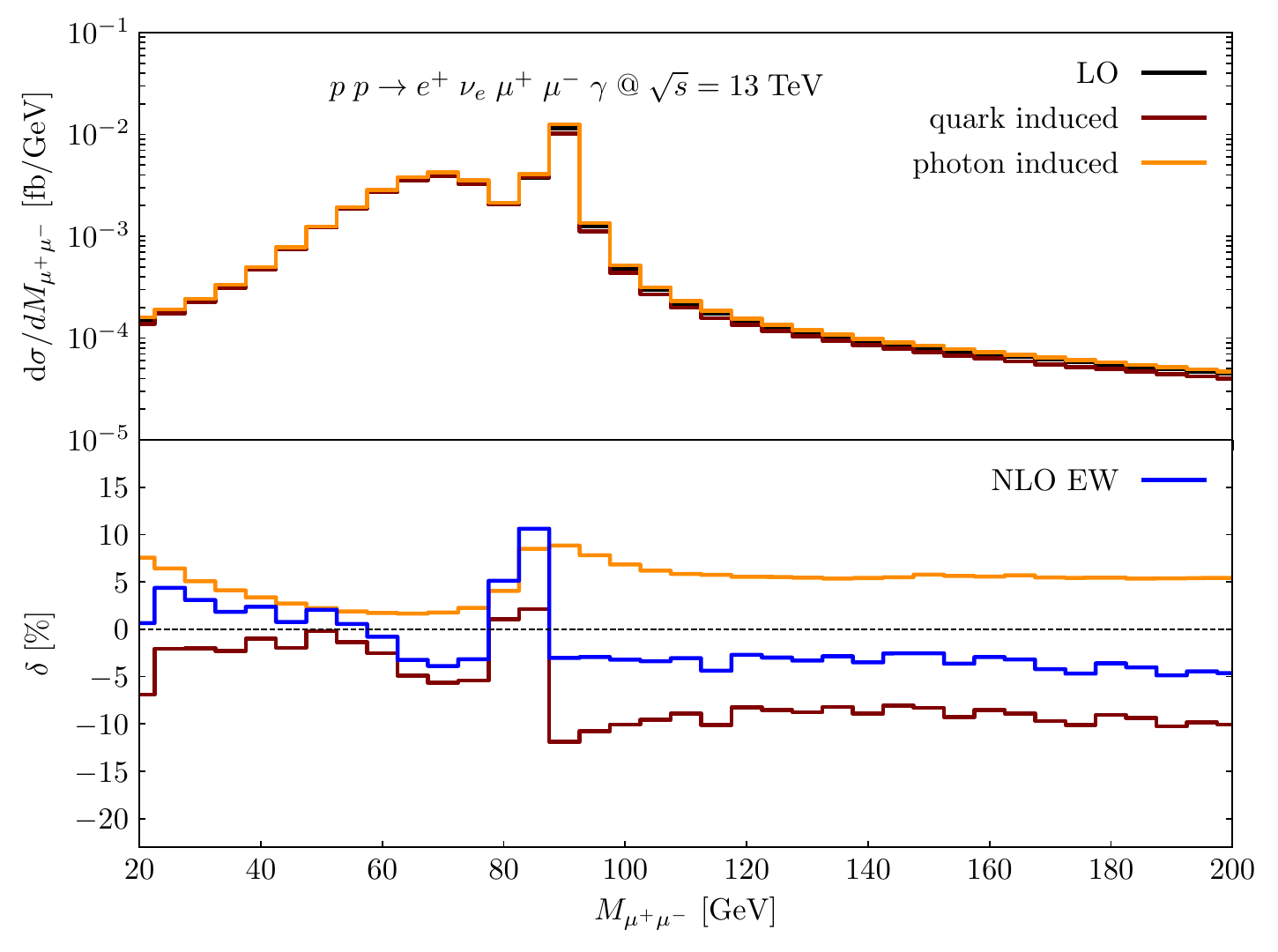}
    \end{minipage}
\hfill
    \begin{minipage}[b]{0.49\textwidth}
    \includegraphics[width=\textwidth]{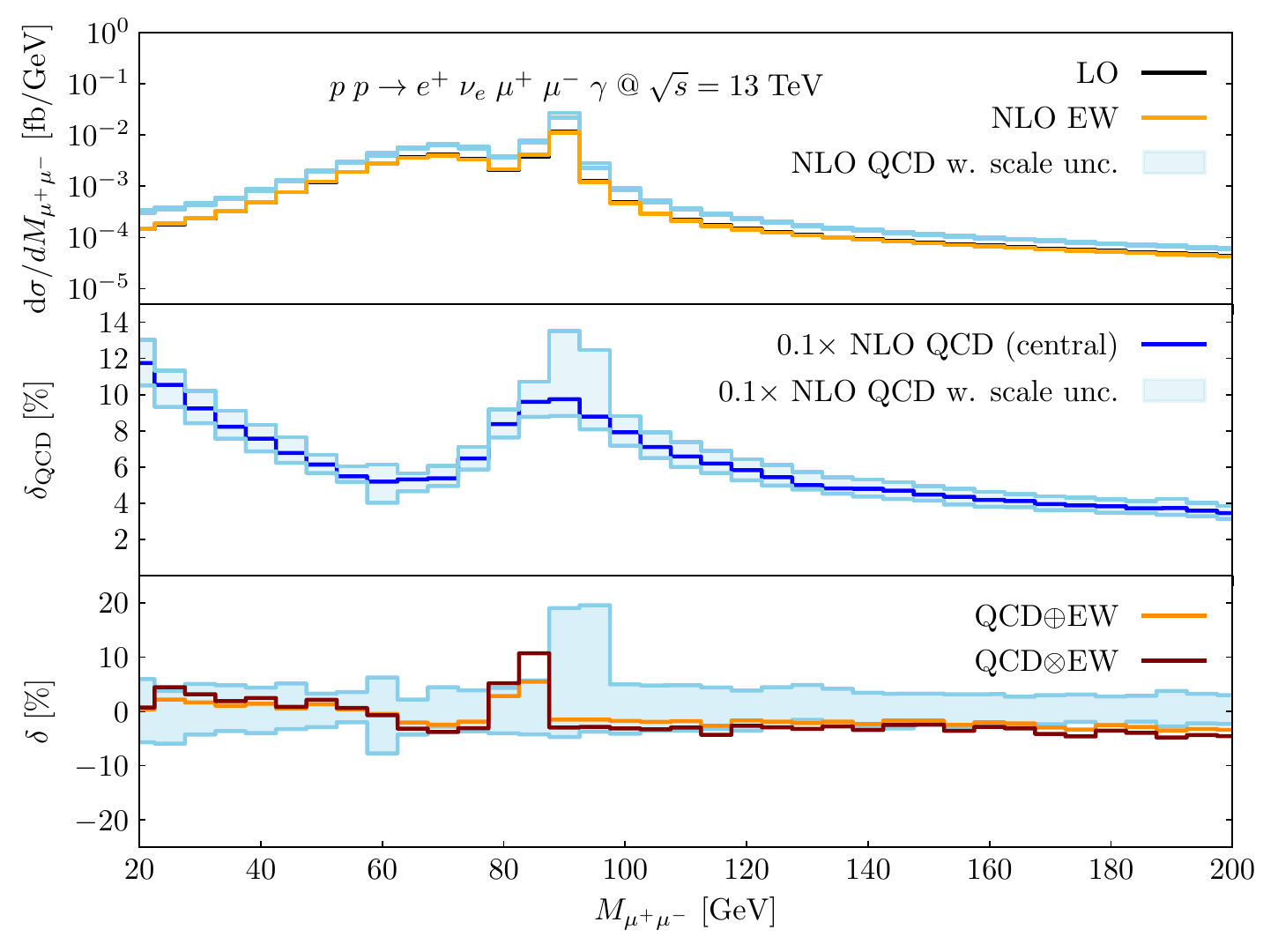}
    \end{minipage}
\caption{LO, NLO QCD, and NLO EW predictions for the distribution in the invariant mass of the $\mu^+\mu^-$ pair in the low-invariant mass region and the corresponding relative corrections. The top panel of the l.h.s plot shows the LO (black), quark-induced (red) and photon-induced (orange) contributions. Their corresponding relative corrections, $\delta=\sigma_{X}/\sigma_{\text{LO}}-1$, with $X$=EW (blue), $q \bar q'$ (red), $\gamma q$ (orange) contributions, are provided on the bottom panel. In top panel of the r.h.s. plot, we display the LO results (black) for the distribution together with NLO EW (orange) and NLO QCD cross sections with QCD scale uncertainties (light blue band). In the middle panel, the relative NLO QCD corrections, $\delta_{\text{QCD}}=\sigma_{\text{QCD}}/\sigma_{\text{LO}}-1$, at the central scale $\mu_{\text{R}}=\mu_{\text{F}}=mu_0$ (blue) along with QCD scale uncertainties (light blue band) are shown, where their sizes in percentage are scaled down by a factor of 10 for a better fit in the plot. In the lower panel, the NLO EW corrections are combined with NLO QCD corrections in the additive (orange) and multiplicative (red) approach, and are displayed together with the NLO QCD uncertainty band (light blue). See the text for a detailed description.}
\label{fig:m_34_nlo_in}
\end{figure}

\begin{figure}[!tbp]
\centering
    \begin{minipage}[b]{0.495\textwidth}
    \includegraphics[width=\textwidth]{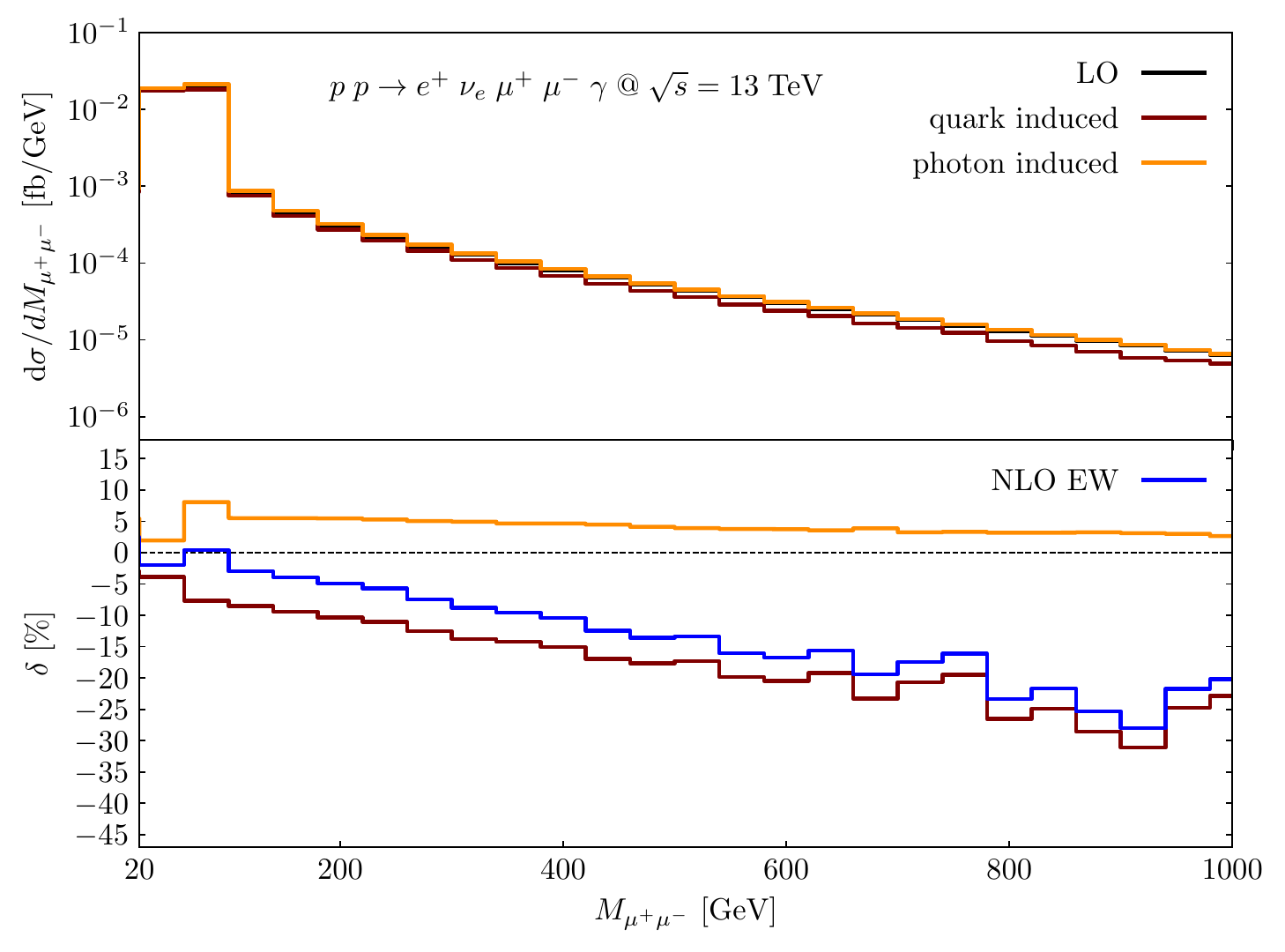}
    \end{minipage}
\hfill
    \begin{minipage}[b]{0.495\textwidth}
    \includegraphics[width=\textwidth]{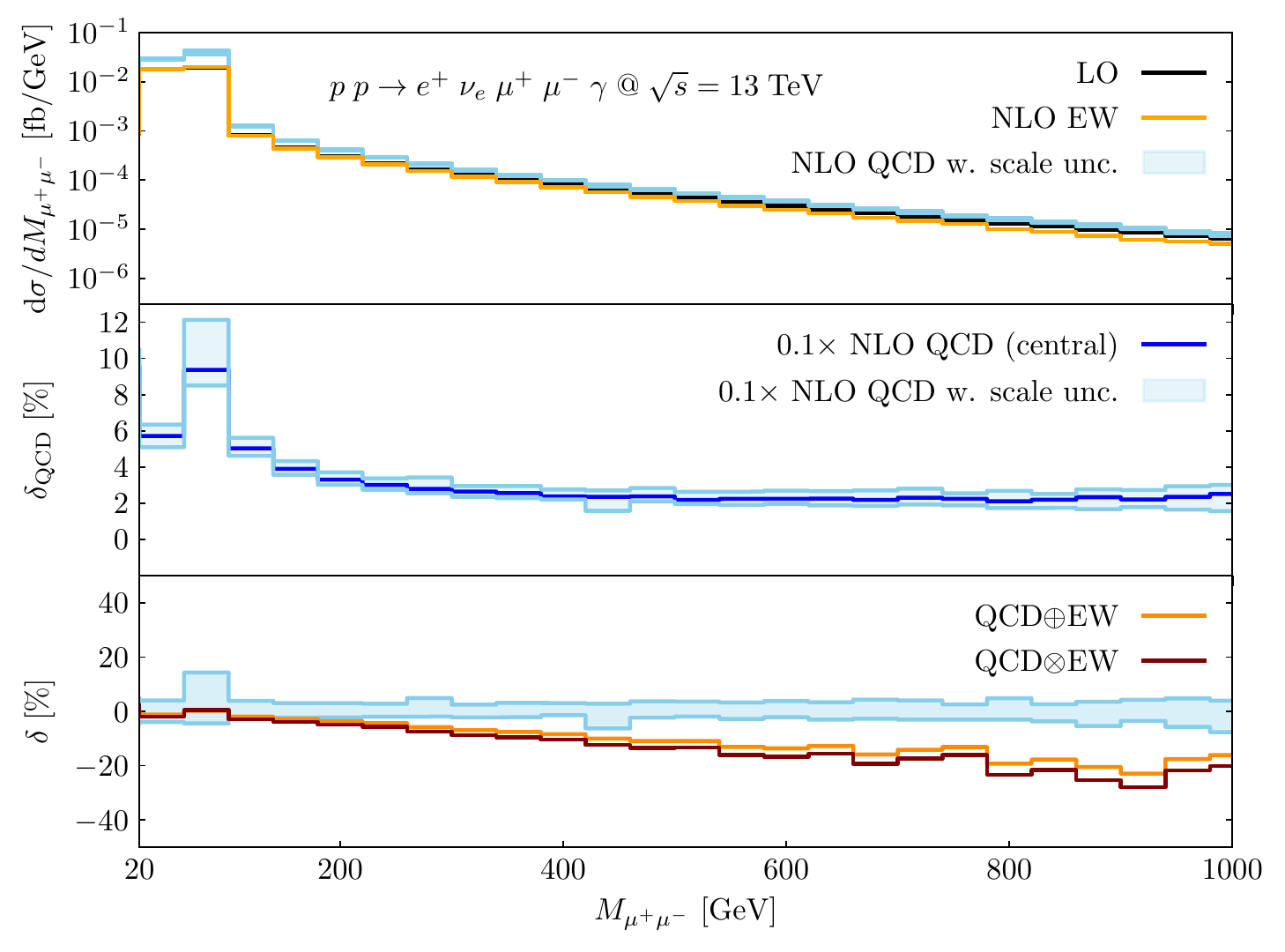}
    \end{minipage}
\caption{LO, NLO QCD, and NLO EW predictions for the distribution in the invariant mass of the $\mu^+\mu^-$ pair in the high-invariant mass region and the corresponding relative corrections with respect to LO predictions. See the caption of FIG.~\ref{fig:m_34_nlo_in} and the text for a detailed description.}
\label{fig:m_34_nlo}
\end{figure}

\begin{figure}[!tbp]
\centering
\begin{minipage}[b]{0.49\textwidth}
    \includegraphics[width=\textwidth]{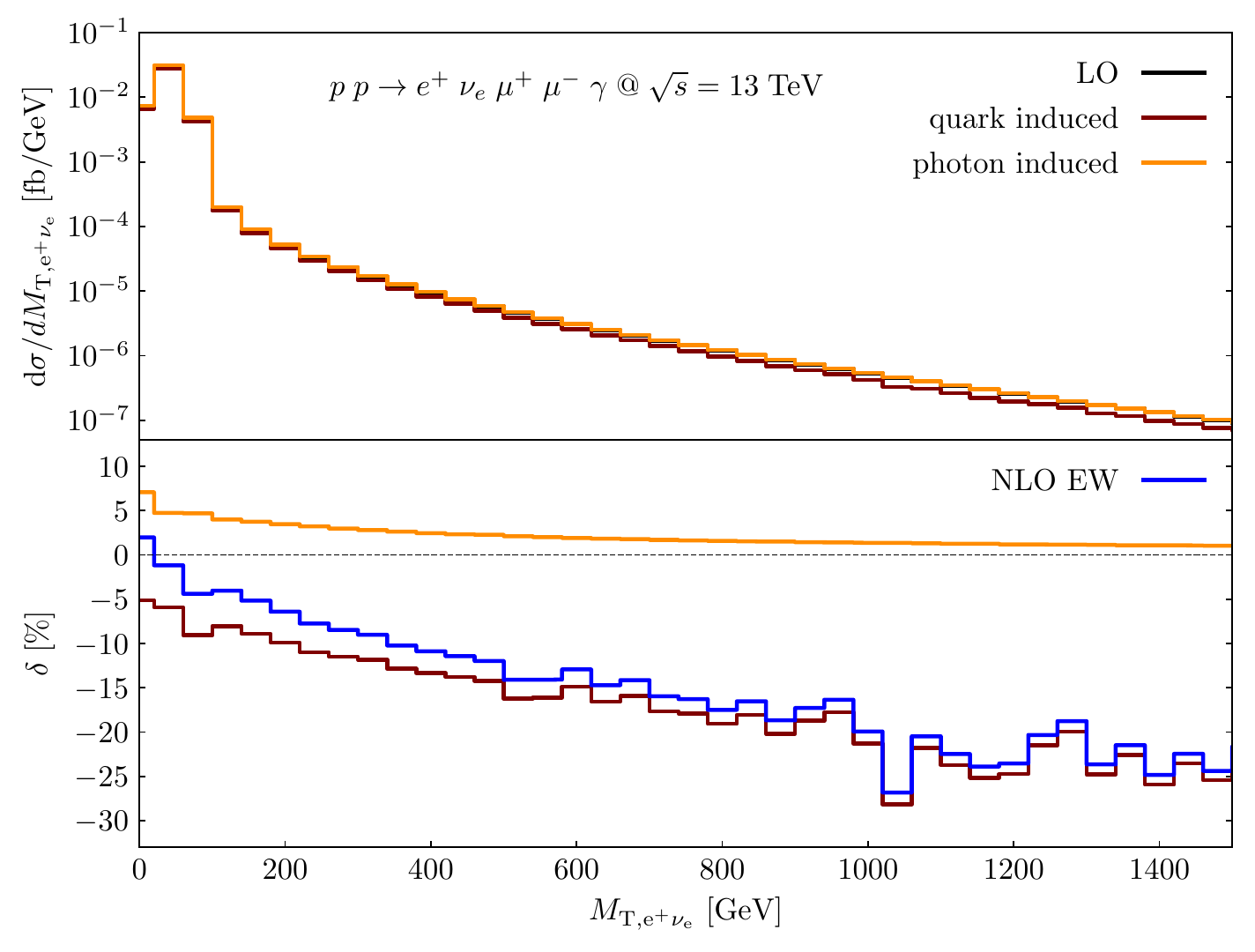}
\end{minipage}
\hfill
\begin{minipage}[b]{0.49\textwidth}
    \includegraphics[width=\textwidth]{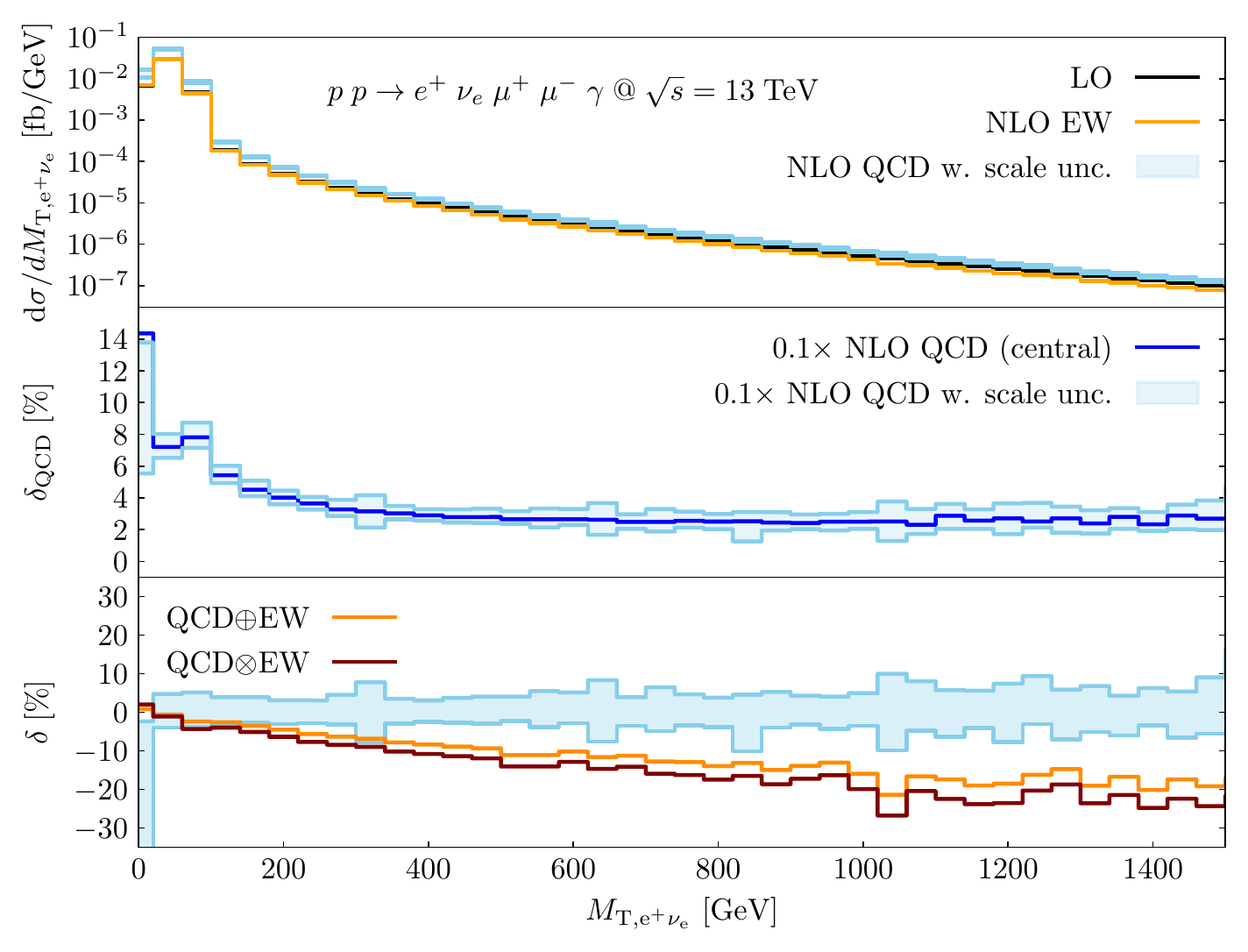}
\end{minipage}
\caption{LO, NLO QCD, and NLO EW predictions for the distribution in the transverse mass of the $e^+ \nu_e$ pair and the corresponding relative corrections. See the caption of FIG.~\ref{fig:m_34_nlo_in} and the text for a detailed description.}
\label{fig:mt_12_nlo}
\end{figure}

\begin{figure}[!tbp]
\centering
\begin{minipage}[b]{0.49\textwidth}
    \includegraphics[width=\textwidth]{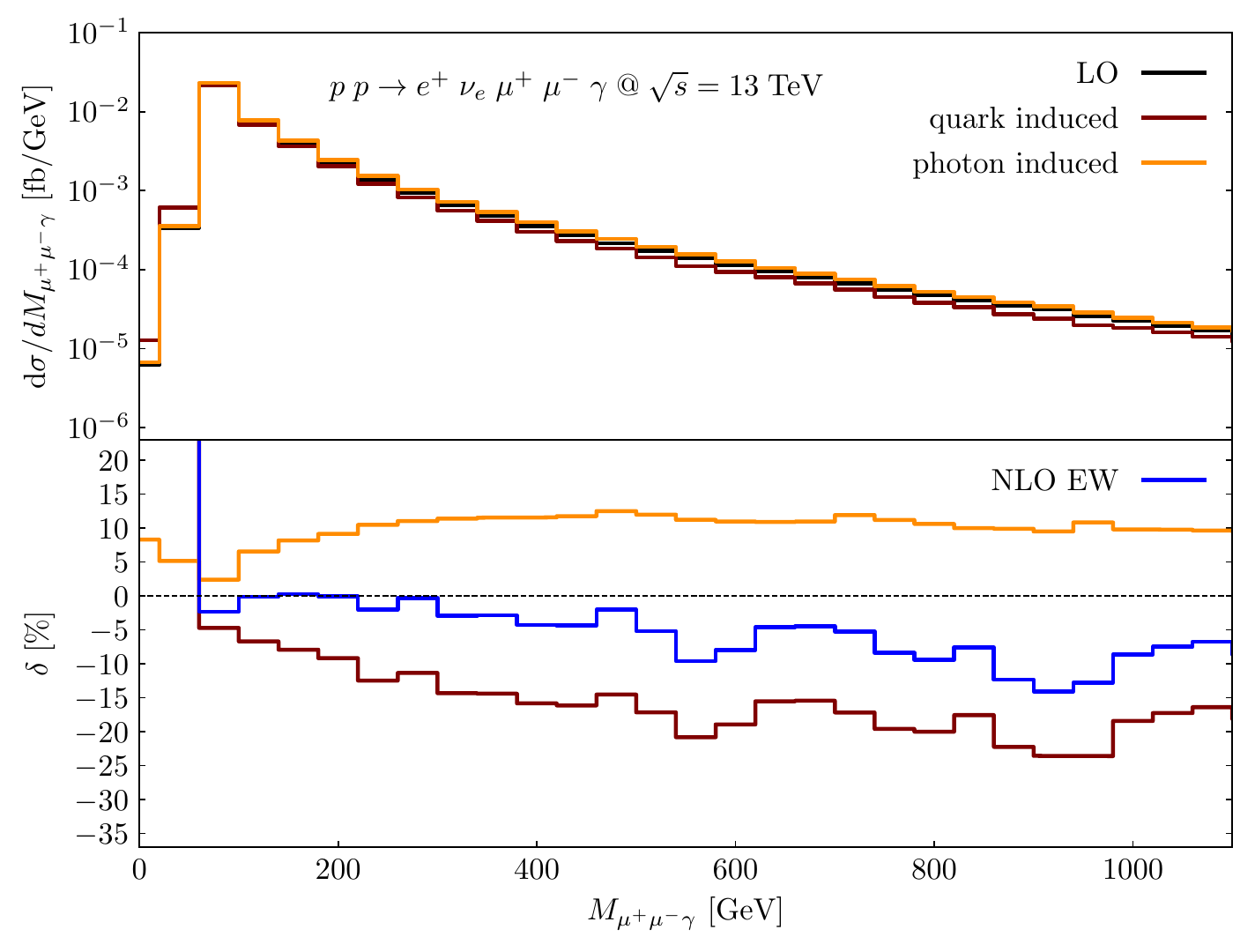}
\end{minipage}
\hfill
\begin{minipage}[b]{0.49\textwidth}
    \includegraphics[width=\textwidth]{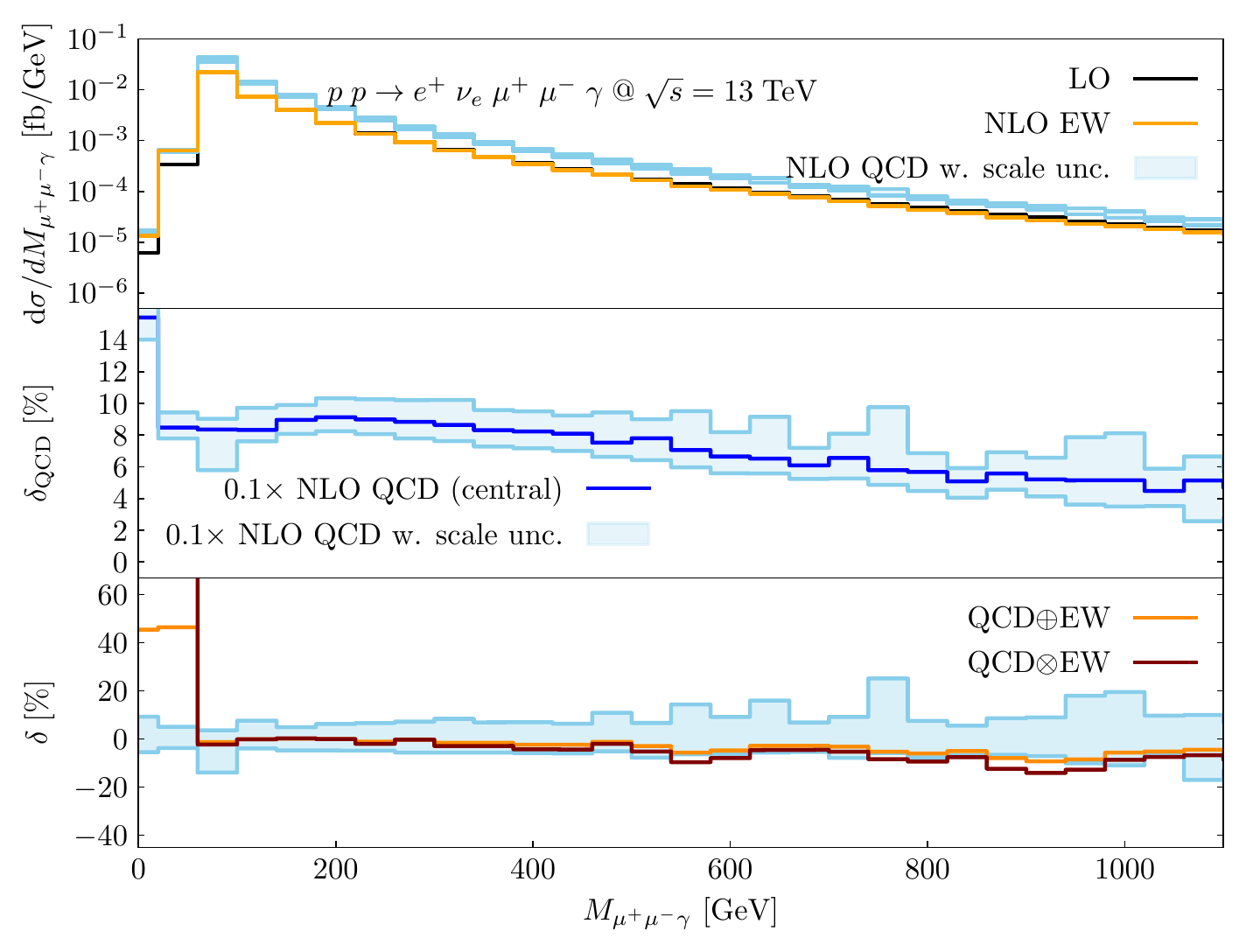}
\end{minipage}
\caption{LO, NLO QCD, and NLO EW predictions for the distribution in the invariant mass of the $\mu^+\mu^-$ pair and the isolated photon, and the corresponding relative corrections. See the caption of FIG.~\ref{fig:m_34_nlo_in} and the text for a detailed description. }
\label{fig:m_345_nlo}
\end{figure}

\begin{figure}[!tbp]
\centering
\begin{minipage}[b]{0.49\textwidth}
    \includegraphics[width=\textwidth]{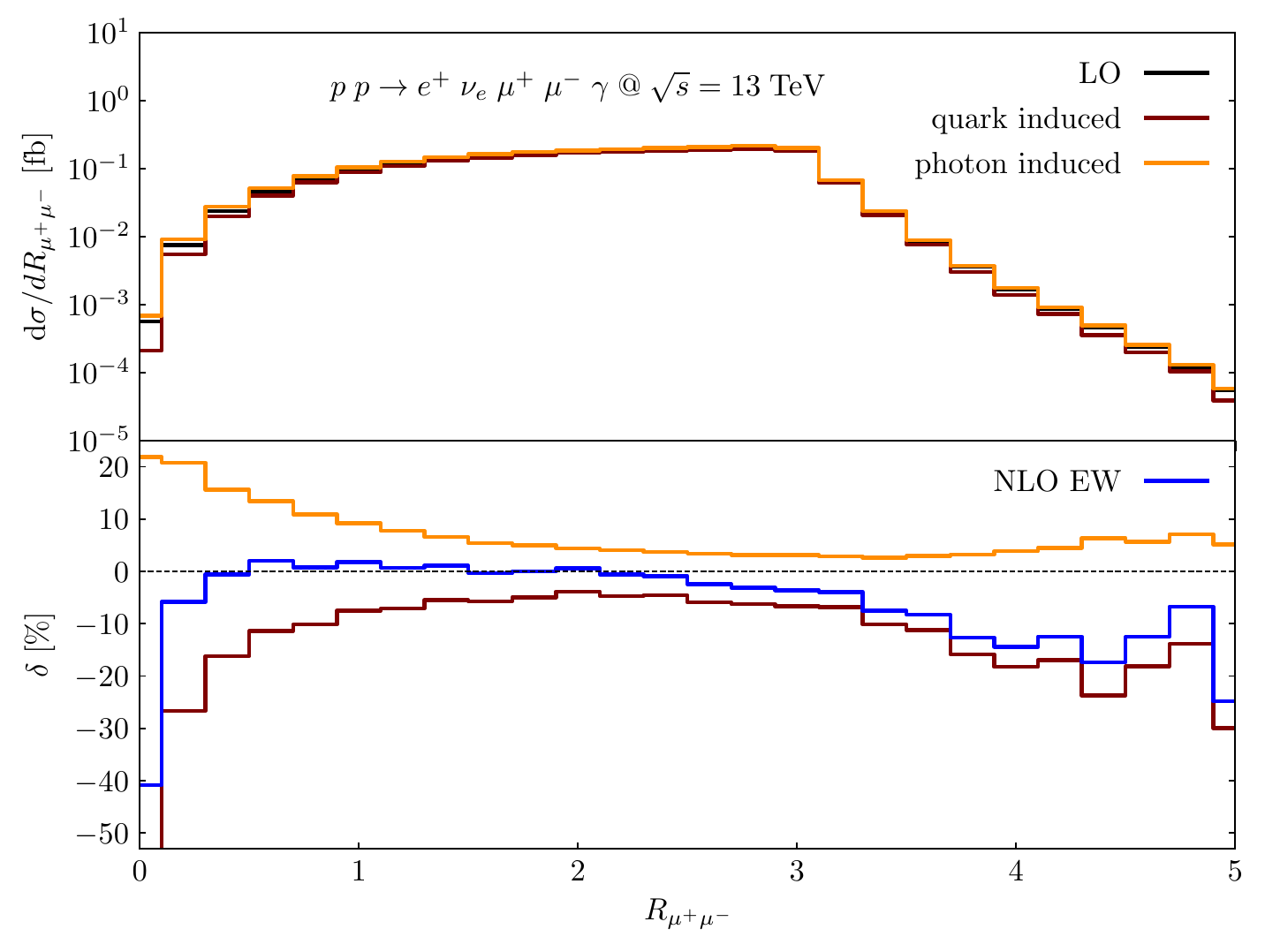}
\end{minipage}
\hfill
\begin{minipage}[b]{0.49\textwidth}
    \includegraphics[width=\textwidth]{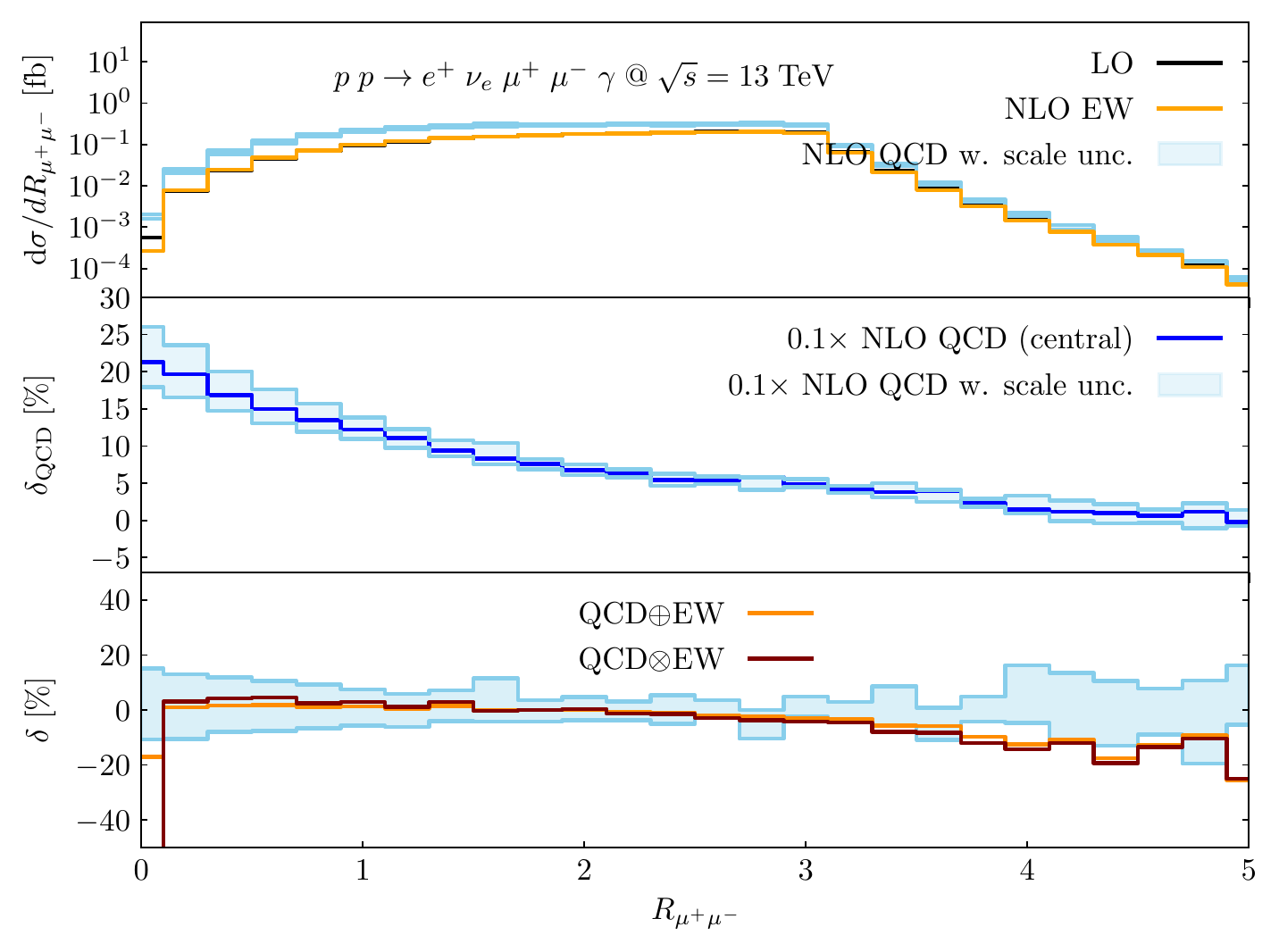}
\end{minipage}
\caption{LO, NLO QCD, and NLO EW predictions for the distribution in the angular separation $R_{ij}$ of Eq.~(\ref{eq: angular separation}) with $i=\mu^+$ and $j=\mu^-$, and the corresponding relative corrections. See the caption of FIG.~\ref{fig:m_34_nlo_in} and the text for a detailed description.}
\label{fig:r_34_nlo}
\end{figure}

\begin{figure}[!tbp]
\centering
\begin{minipage}[b]{0.49\textwidth}
    \includegraphics[width=\textwidth]{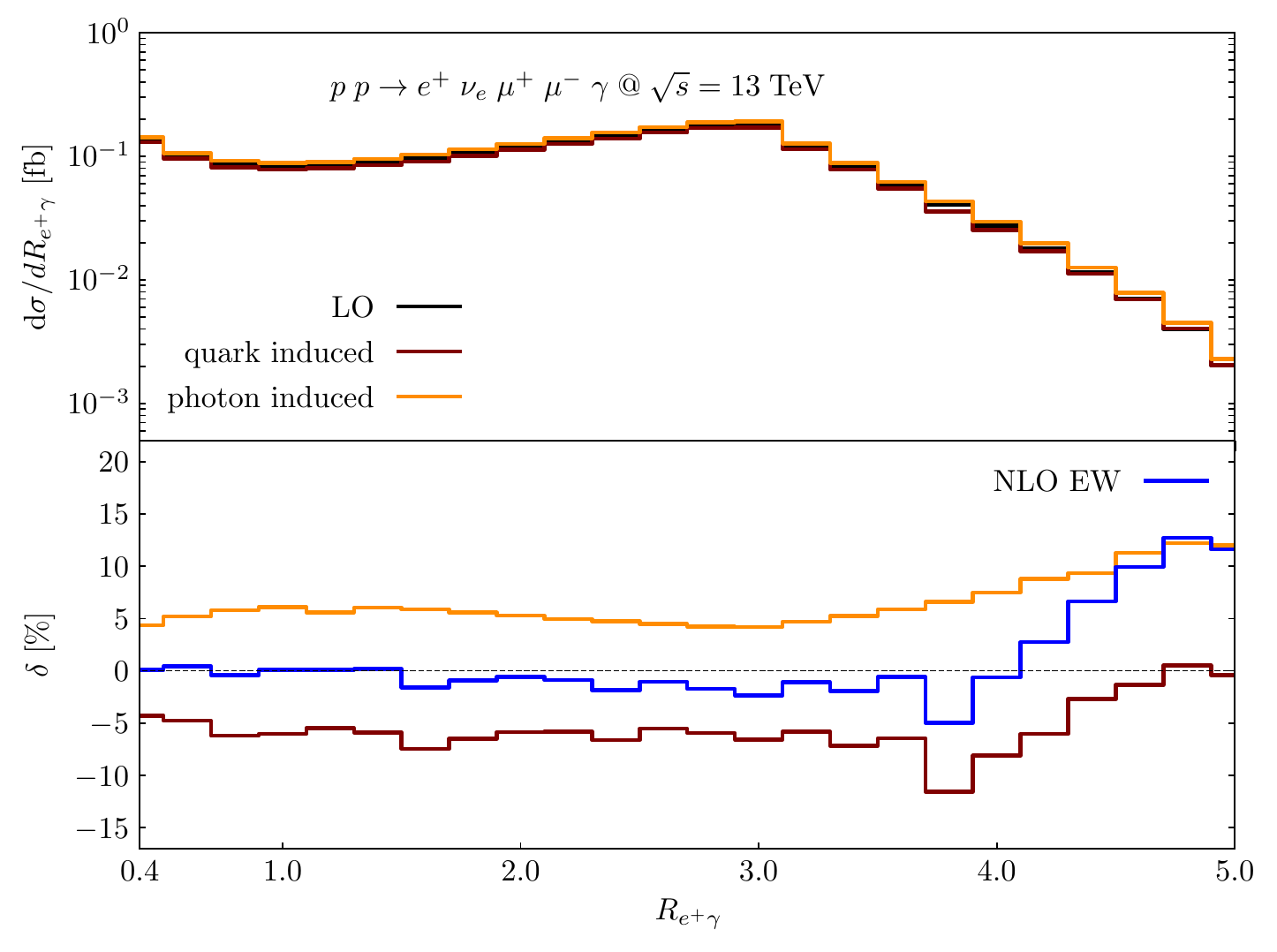}
\end{minipage}
\hfill
\begin{minipage}[b]{0.49\textwidth}
    \includegraphics[width=\textwidth]{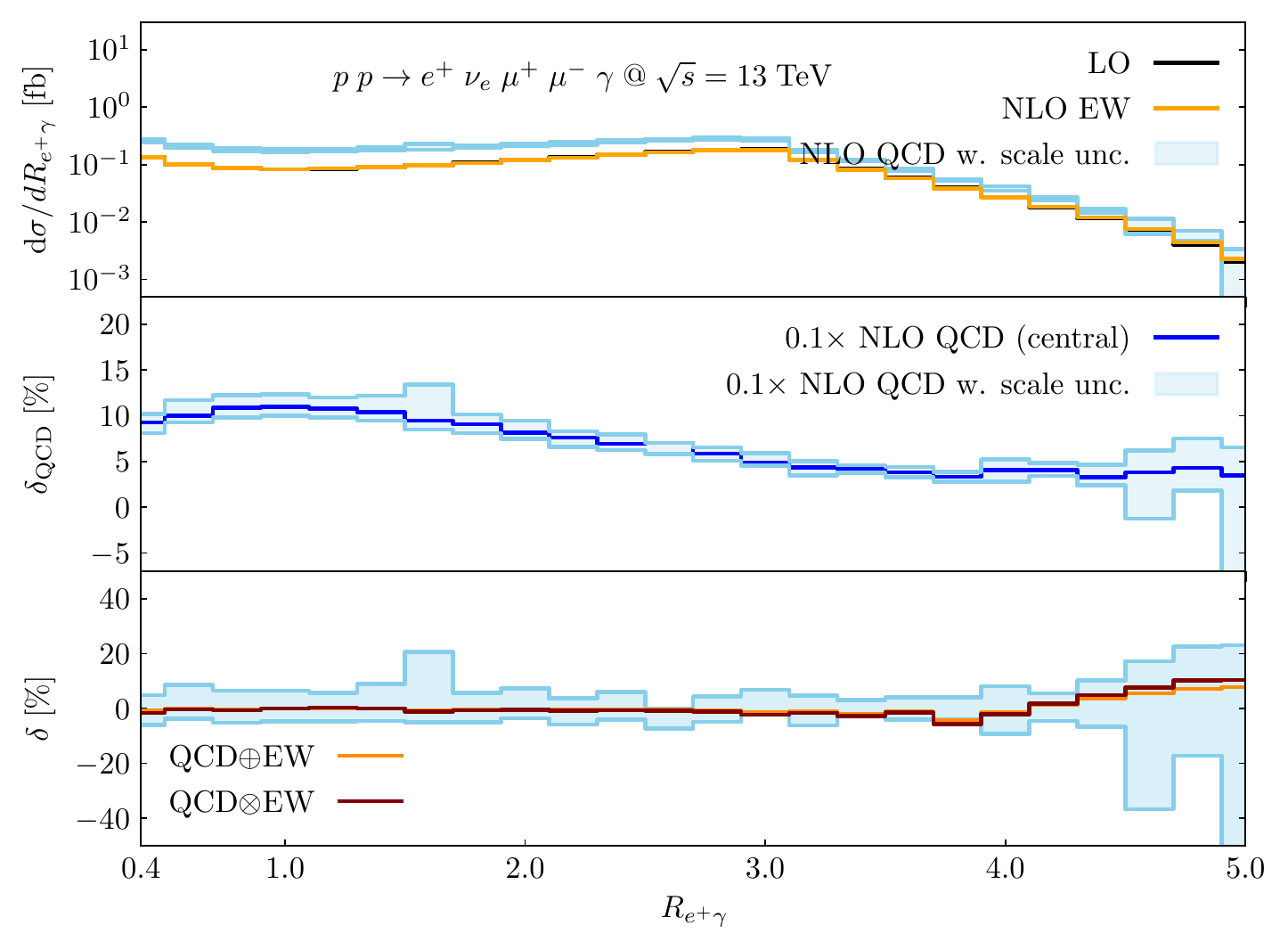}
\end{minipage}
\caption{
LO, NLO QCD, and NLO EW predictions for the distribution in the angular separation $R_{ij}$ of Eq.~(\ref{eq: angular separation}) with $i=e^+$ and $j=\gamma$, and the corresponding relative corrections. See the caption of FIG.~\ref{fig:m_34_nlo_in} and the text for a detailed description.}
\label{fig:r_15_nlo}
\end{figure}

\begin{figure}[!tbp]
\centering
\begin{minipage}[b]{0.49\textwidth}
    \includegraphics[width=\textwidth]{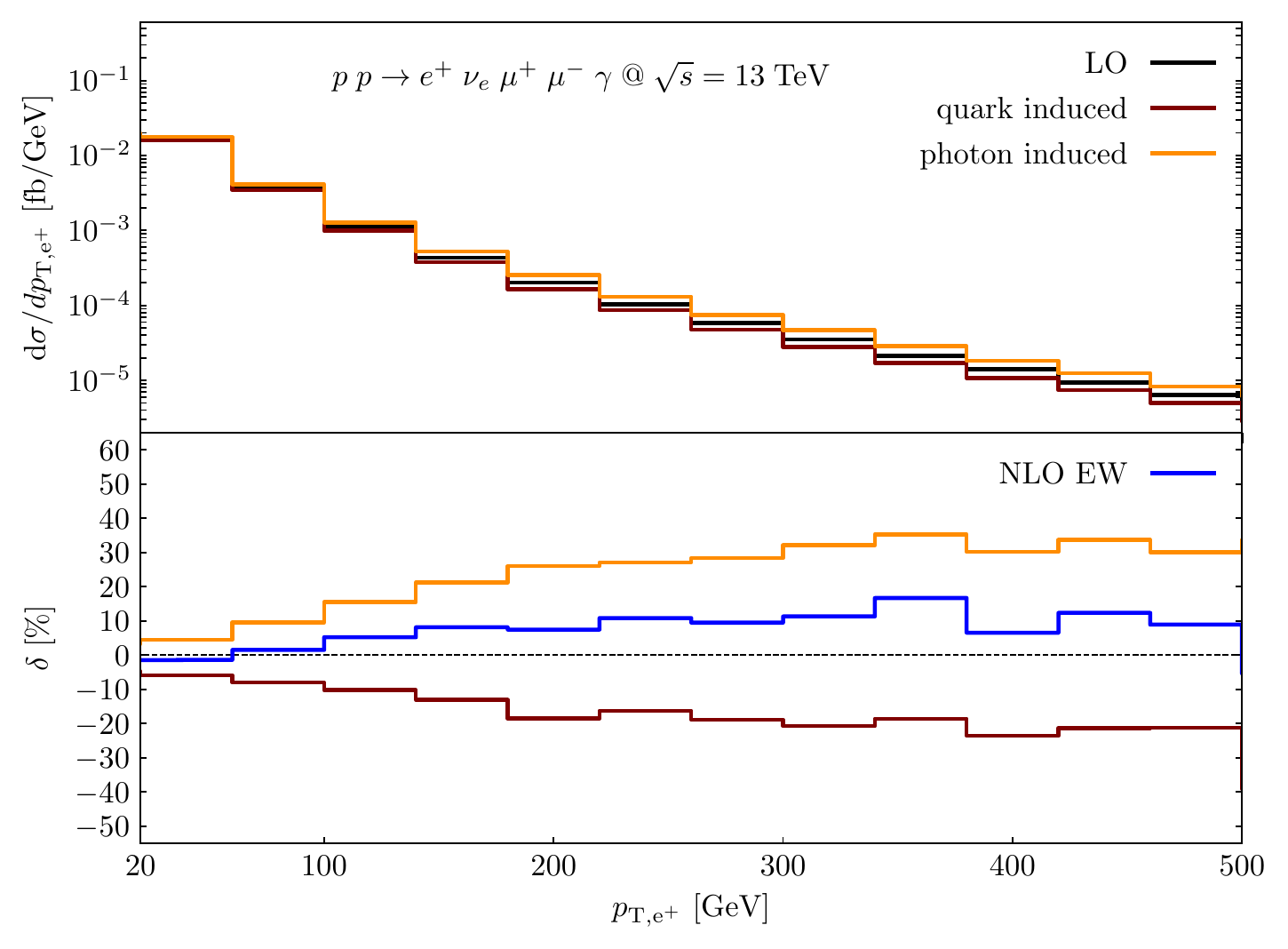}
\end{minipage}
\hfill
\begin{minipage}[b]{0.49\textwidth}
    \includegraphics[width=\textwidth]{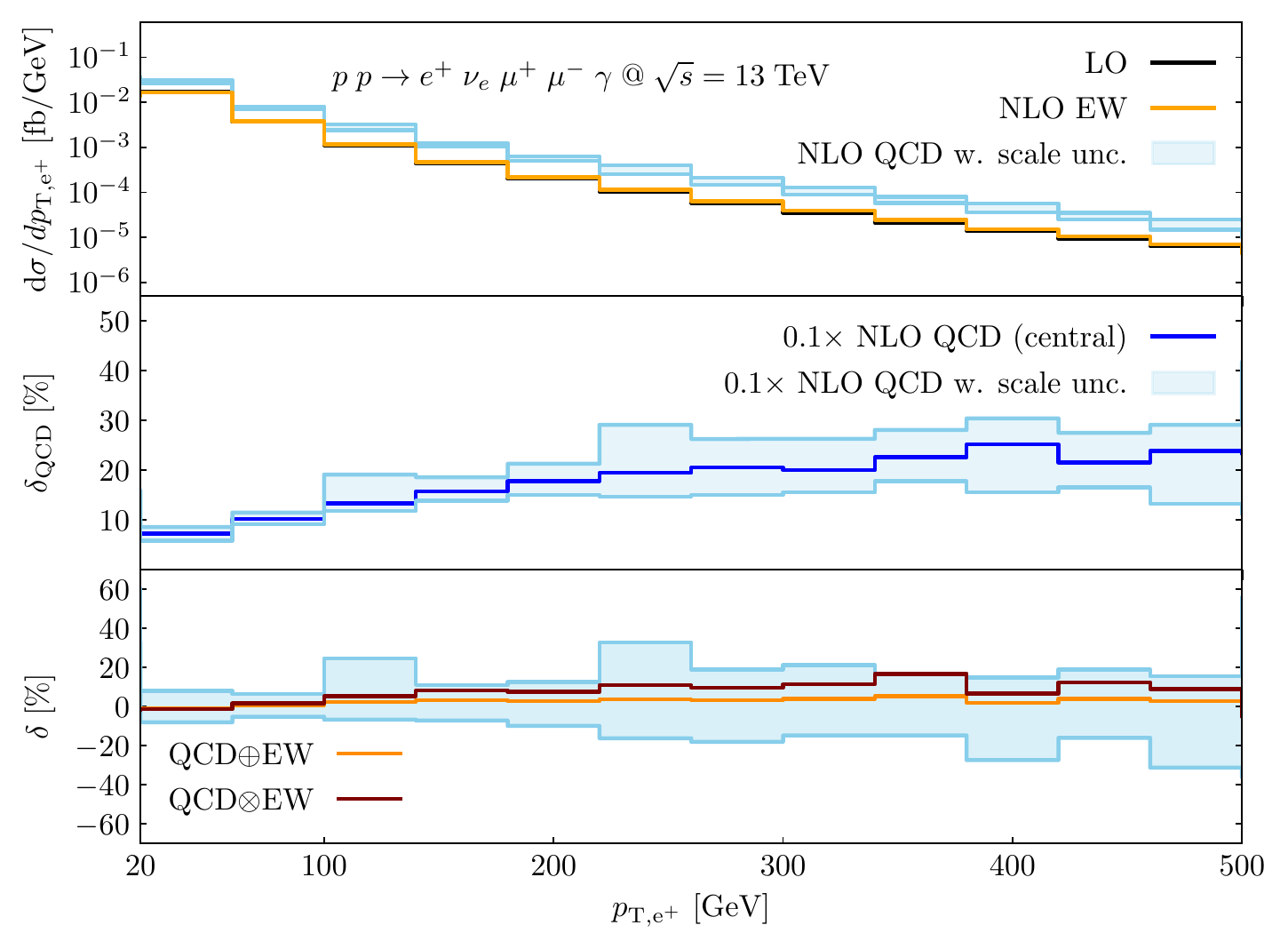}
\end{minipage}
\caption{LO, NLO QCD, and NLO EW predictions for the distribution in the transverse momentum of the positron, and the corresponding relative corrections. See the caption of FIG.~\ref{fig:m_34_nlo_in} and the text for a detailed description.}
\label{fig:pt_1_nlo}
\end{figure}

\begin{figure}[!tbp]
\centering
\begin{minipage}[b]{0.49\textwidth}
    \includegraphics[width=\textwidth]{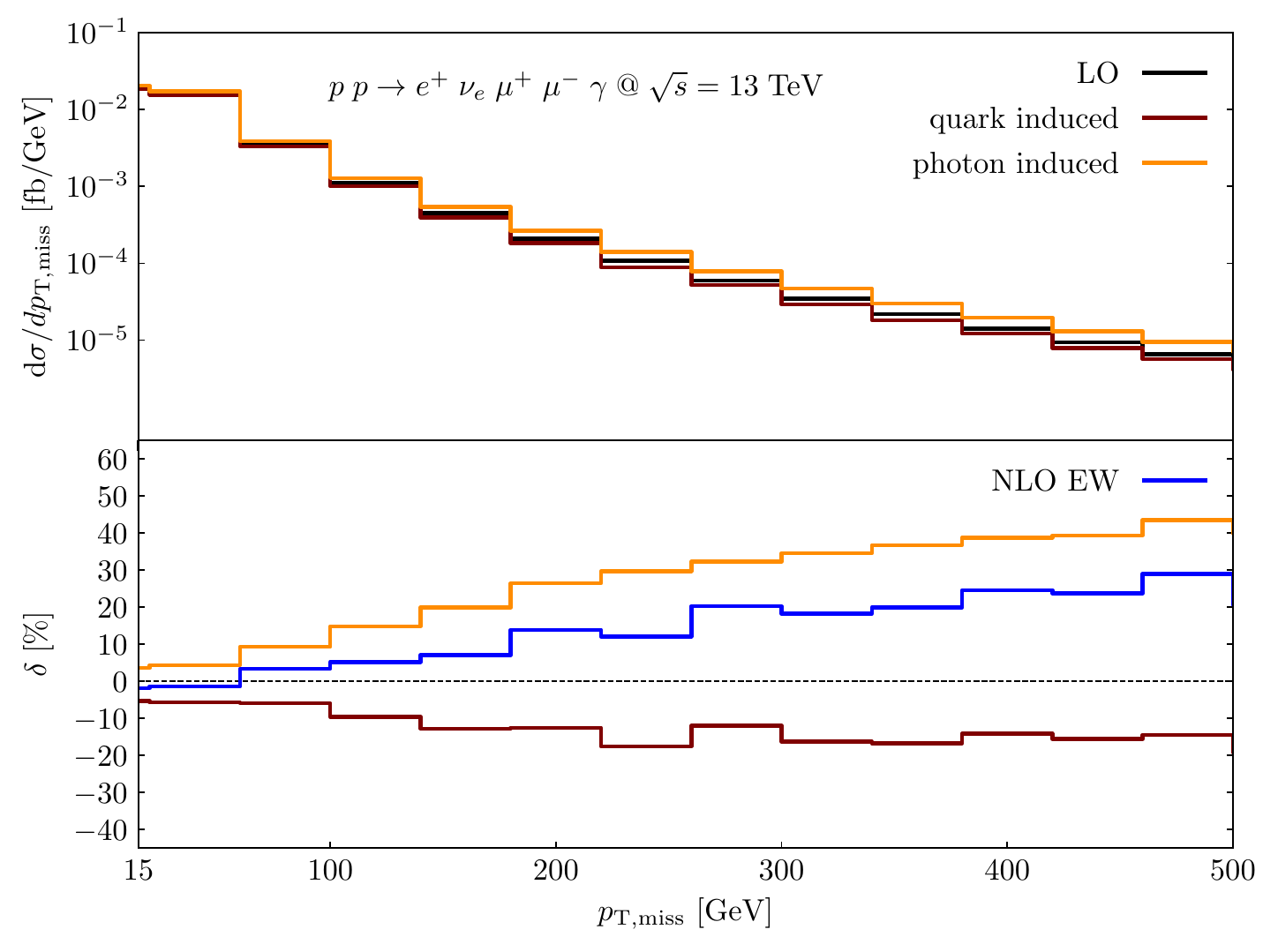}
\end{minipage}
\hfill
\begin{minipage}[b]{0.49\textwidth}
    \includegraphics[width=\textwidth]{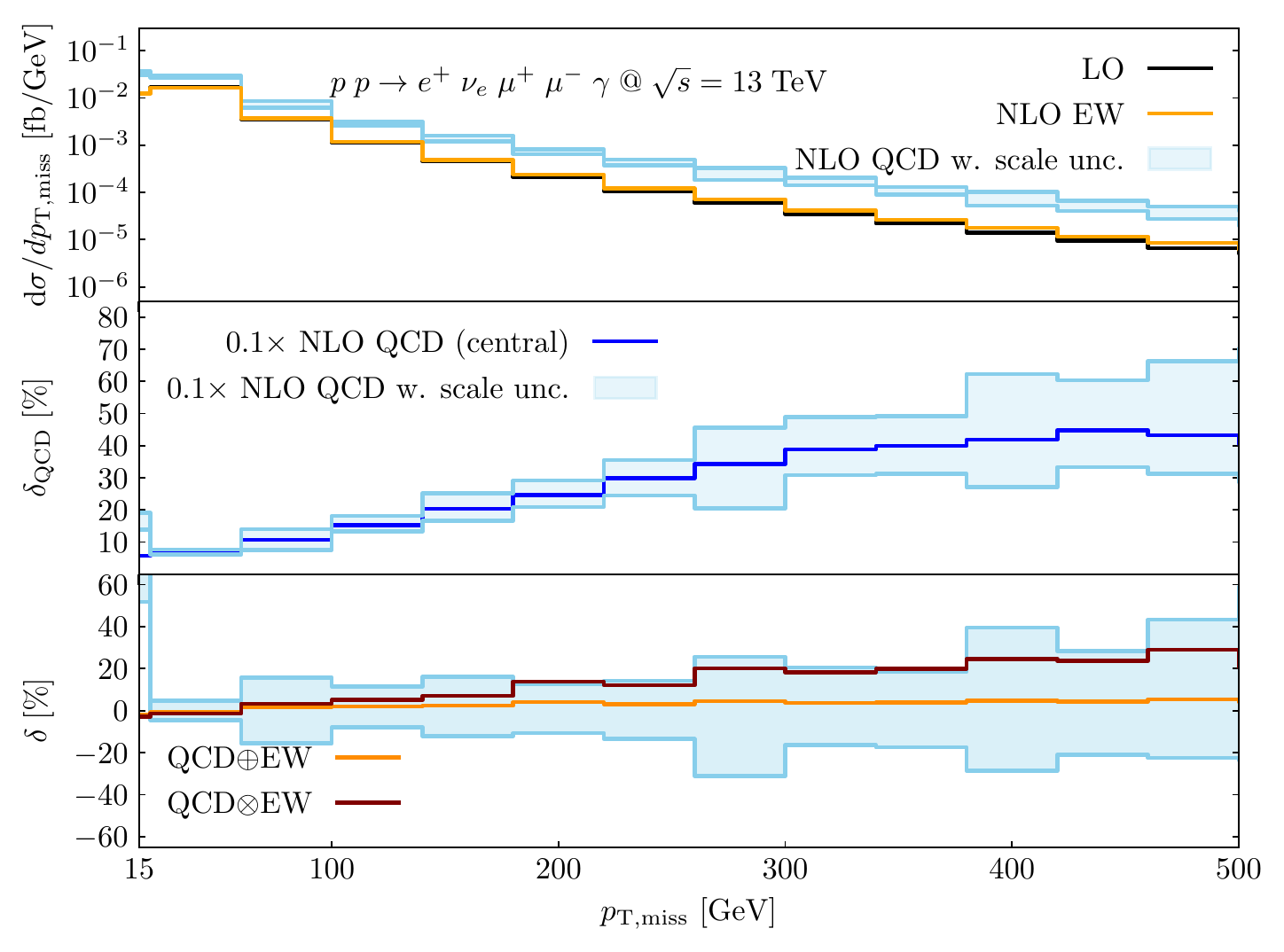}
\end{minipage}
\caption{LO, NLO QCD, and NLO EW predictions for the distribution in the missing transverse momentum, and the corresponding relative corrections. See the caption of FIG.~\ref{fig:m_34_nlo_in} and the text for a detailed description.}
\label{fig:pt_miss_nlo}
\end{figure}

\begin{figure}[!tbp]
\centering
\begin{minipage}[b]{0.49\textwidth}
    \includegraphics[width=\textwidth]{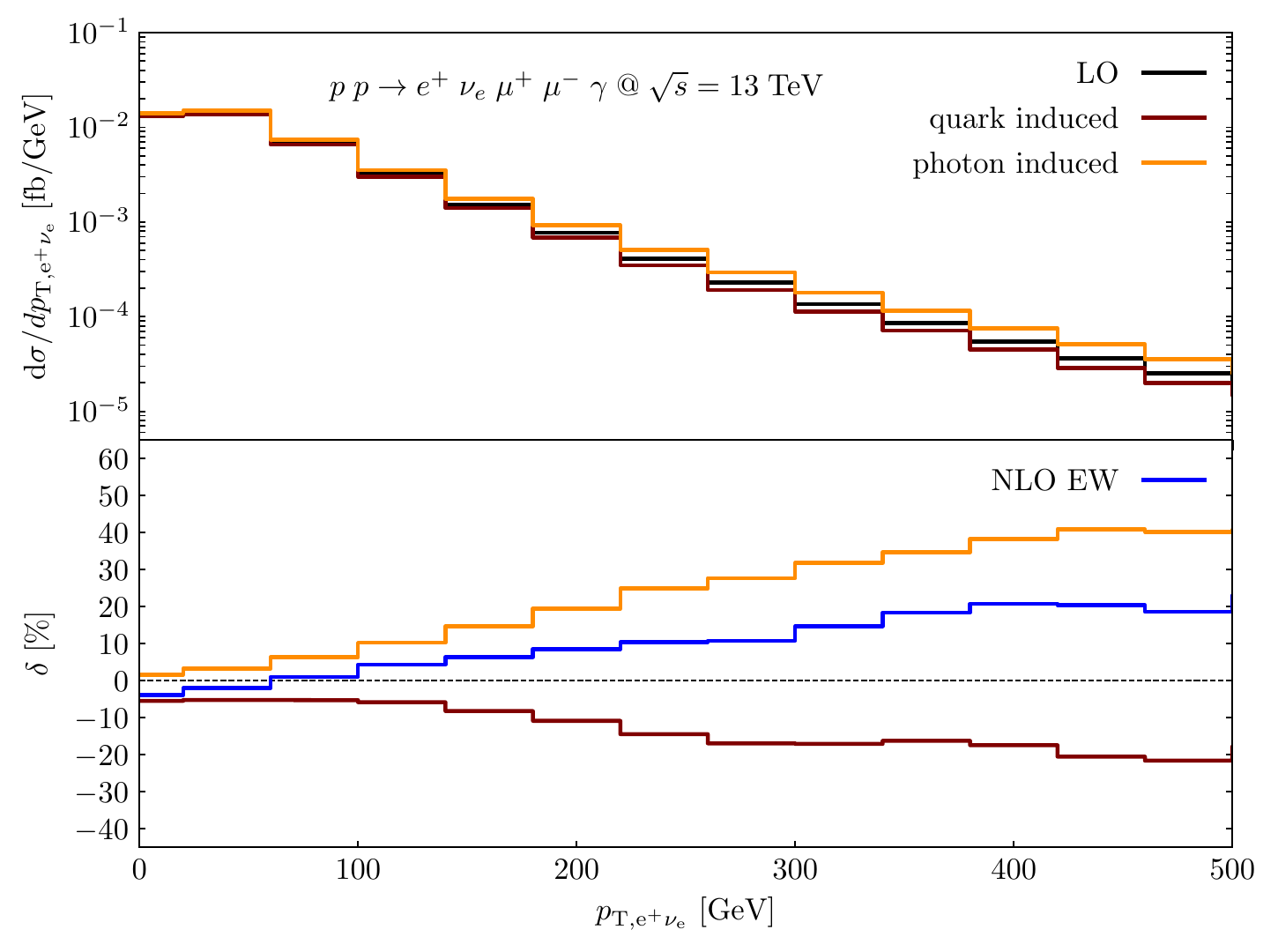}
\end{minipage}
\hfill
\begin{minipage}[b]{0.49\textwidth}
    \includegraphics[width=\textwidth]{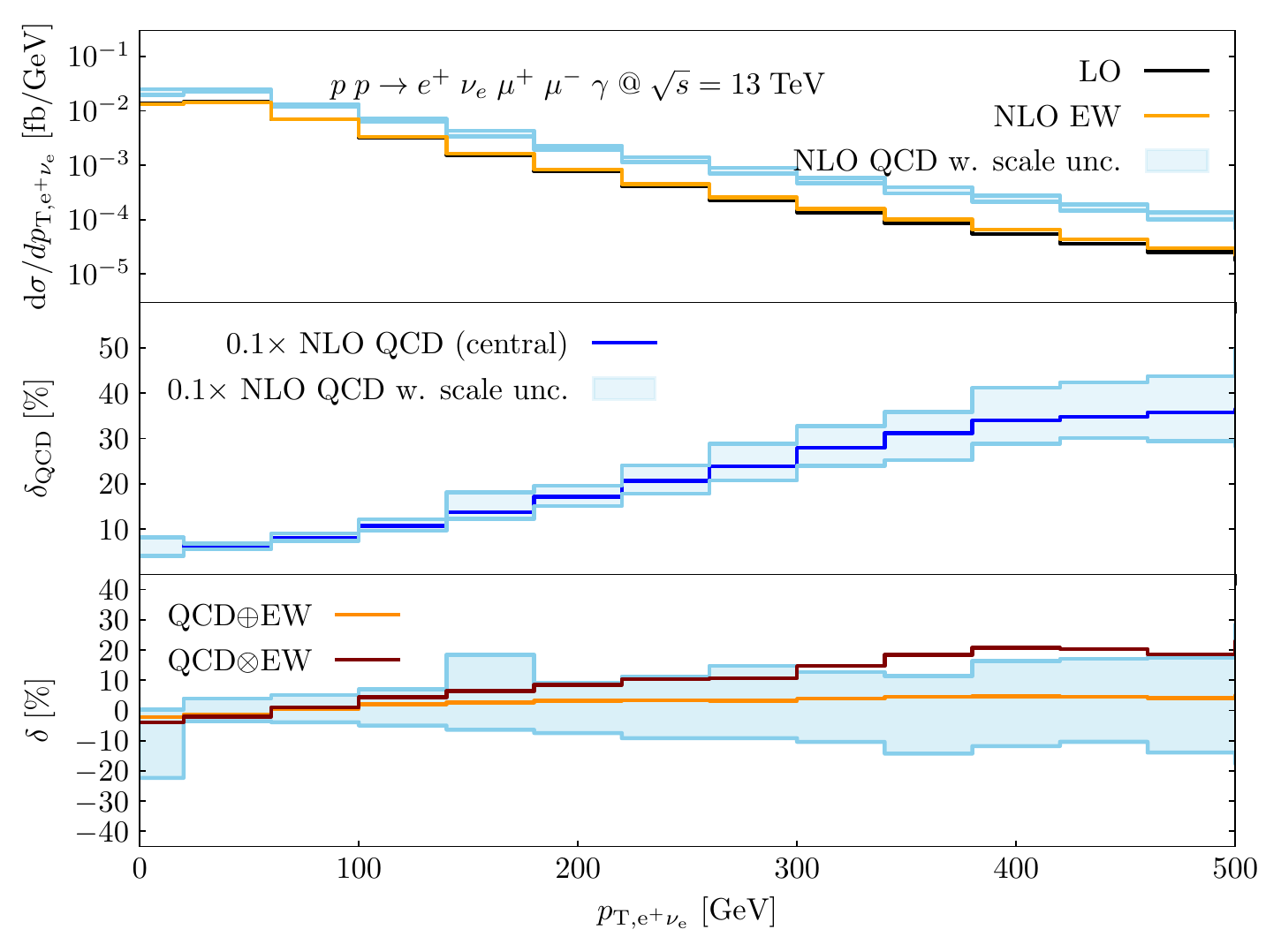}
\end{minipage}
\caption{LO, NLO QCD, and NLO EW predictions for the distribution in the transverse momentum of the $e^+\nu_e$ pair, and the corresponding relative corrections. See the caption of FIG.~\ref{fig:m_34_nlo_in} and the text for a detailed description.}
\label{fig:pt_w_nlo}
\end{figure}

\begin{figure}[!tbp]
\centering
\begin{minipage}[b]{0.49\textwidth}
    \includegraphics[width=\textwidth]{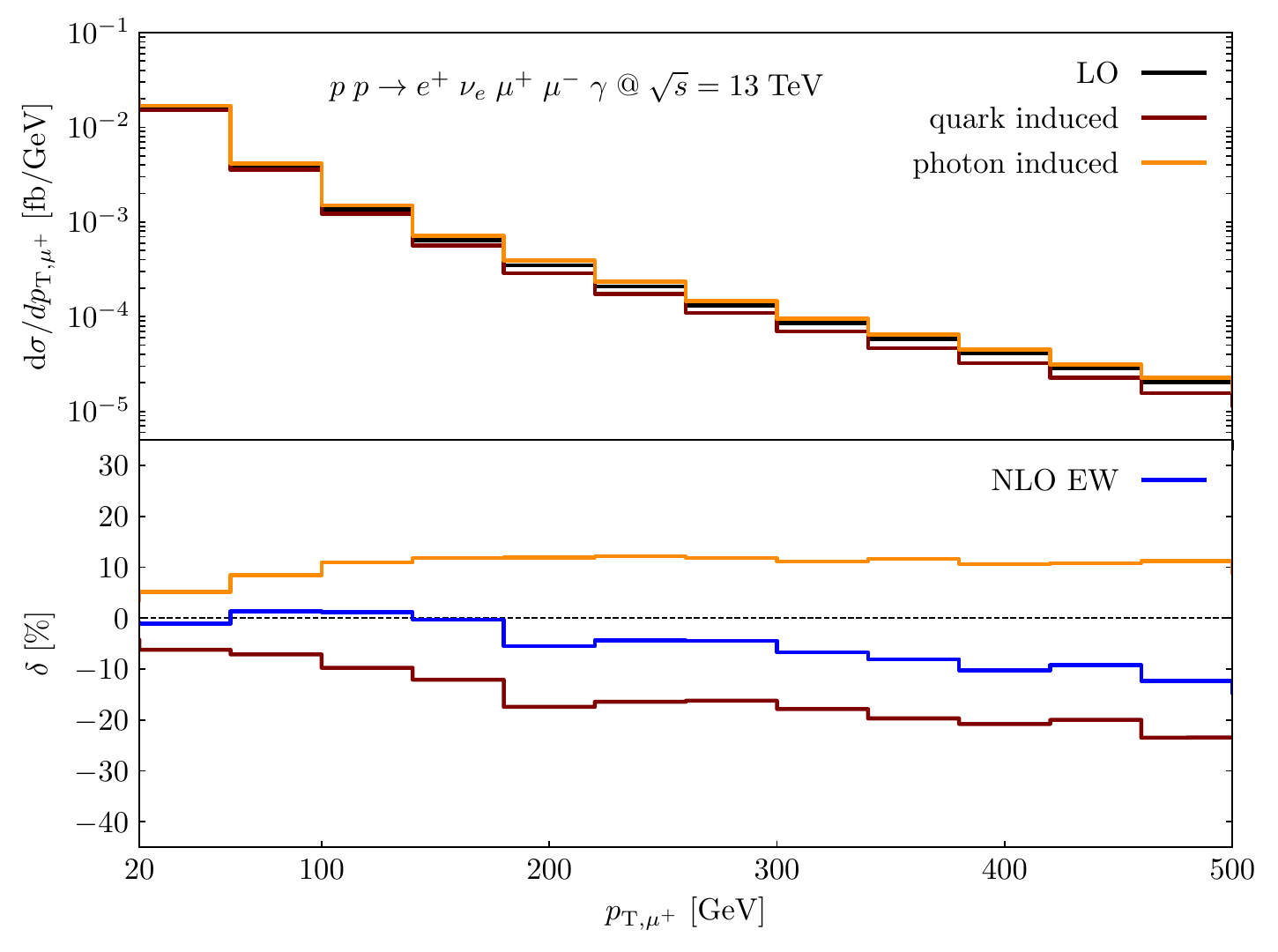}
\end{minipage}
\hfill
\begin{minipage}[b]{0.49\textwidth}
    \includegraphics[width=\textwidth]{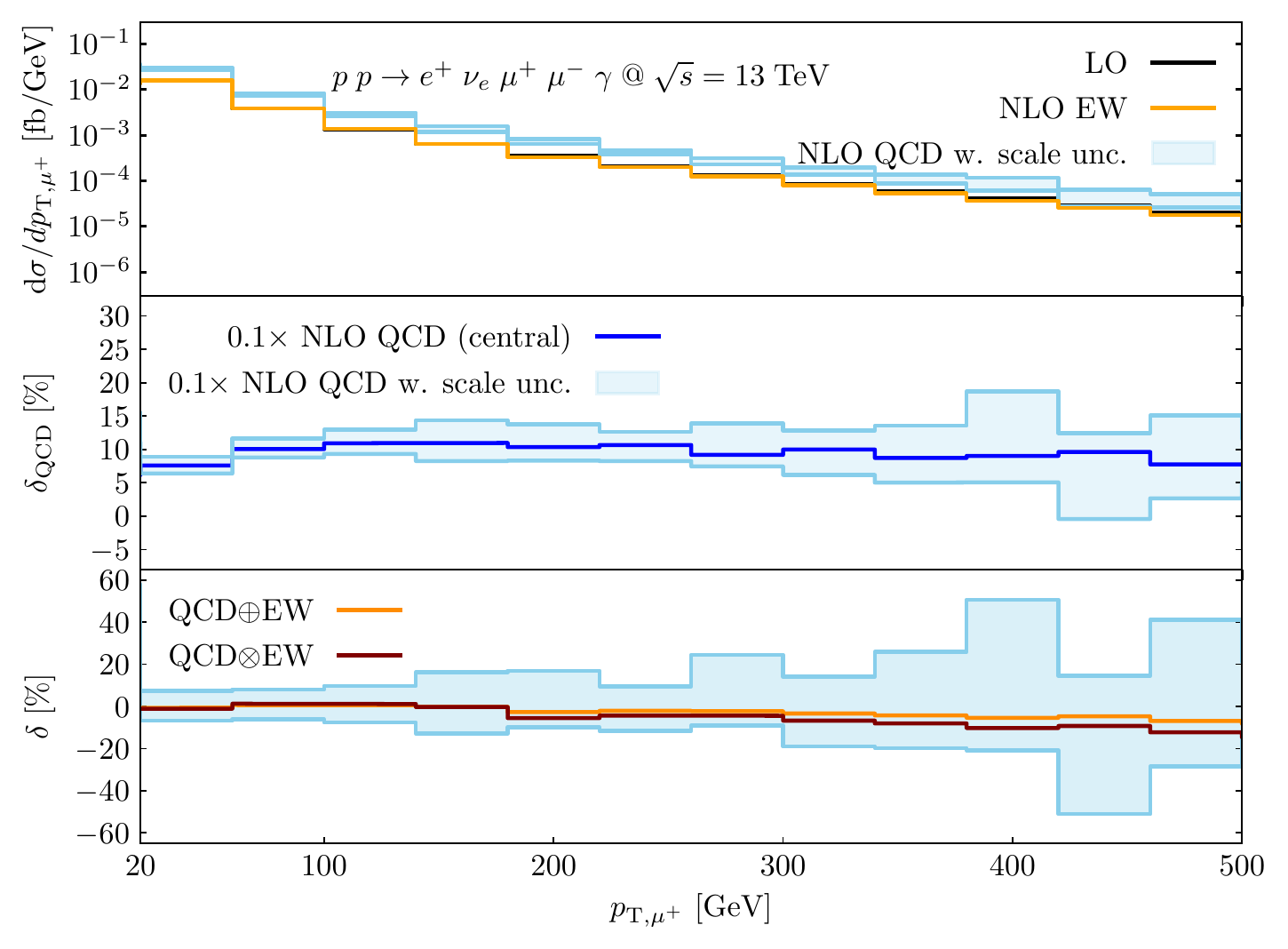}
\end{minipage}
\caption{LO, NLO QCD, and NLO EW predictions for the distribution in the transverse momentum of the $\mu^+$, and the corresponding relative corrections. See the caption of FIG.~\ref{fig:m_34_nlo_in} and the text for a detailed description.}
\label{fig:pt_3_nlo}
\end{figure}

\begin{figure}[!tbp]
\centering
\begin{minipage}[b]{0.49\textwidth}
    \includegraphics[width=\textwidth]{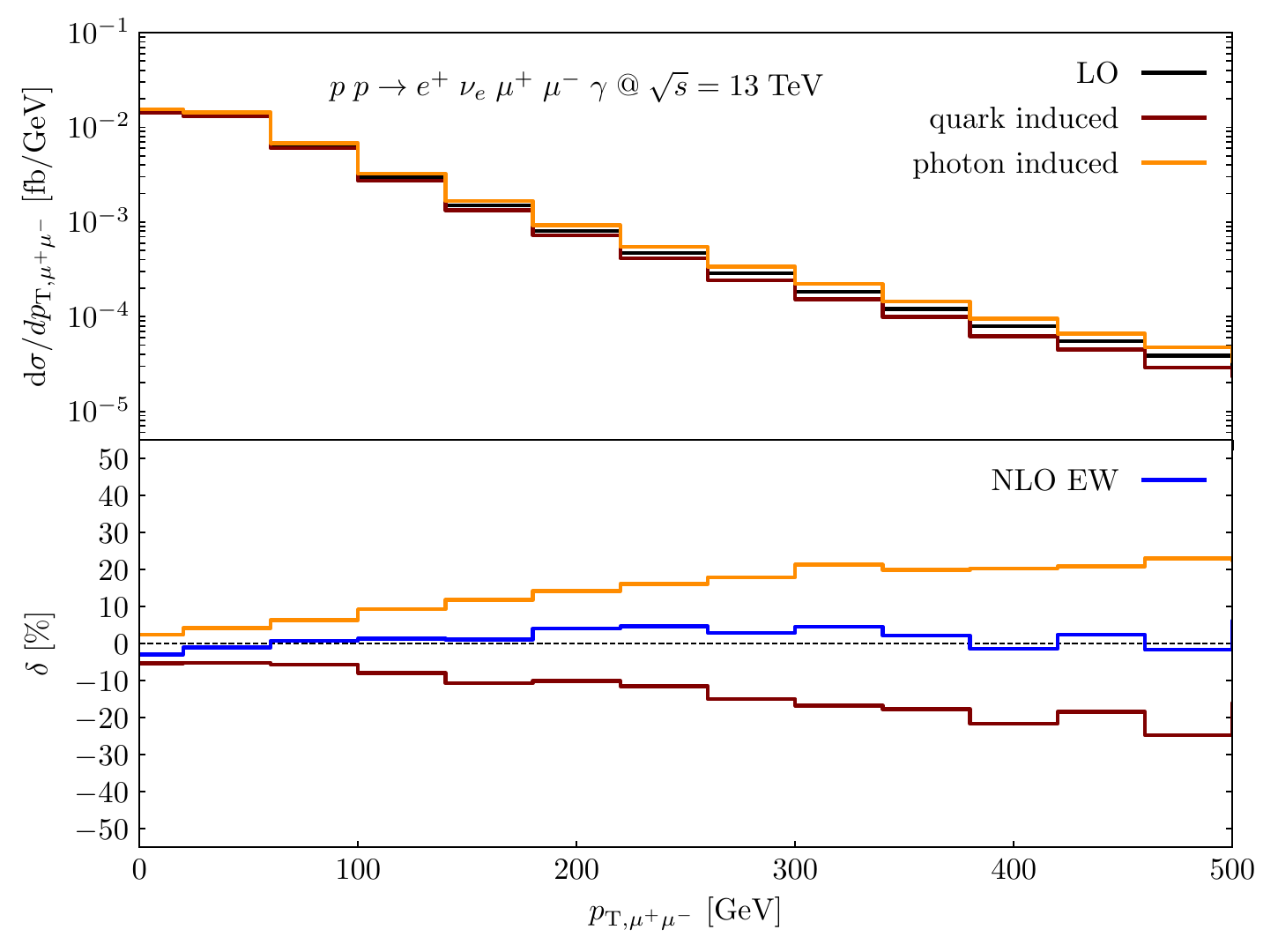}
\end{minipage}
\hfill
\begin{minipage}[b]{0.49\textwidth}
    \includegraphics[width=\textwidth]{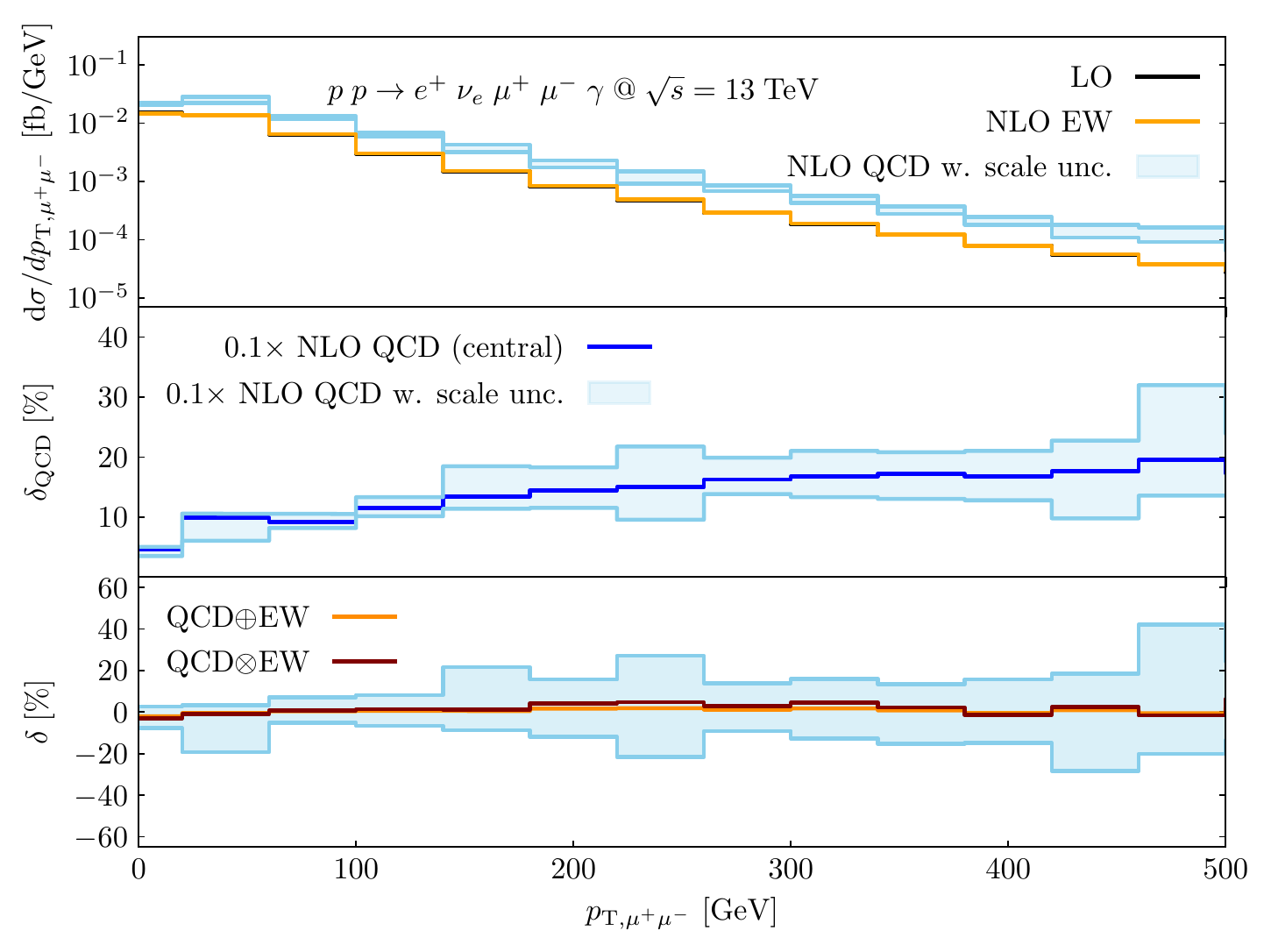}
\end{minipage}
\caption{LO, NLO QCD, and NLO EW predictions for the distribution in the transverse momentum of the $\mu^+\mu^-$ pair, and the corresponding relative corrections. See the caption of FIG.~\ref{fig:m_34_nlo_in} and the text for a detailed description.}
\label{fig:pt_z_nlo}
\end{figure}

Despite of the smallness of NLO EW corrections at the total cross section level compared to the large NLO QCD corrections, their impact can be significant in certain differential cross sections and kinematic regions. In order to illustrate the effects of NLO EW corrections, we show in FIGs.~\ref{fig:m_34_nlo_in} - \ref{fig:pt_z_nlo} a pair of plots with results for each distribution as follows: on the left-hand-side plot of each figure, we display the quark-induced (red) and photon-induced (orange) contributions to the NLO EW distribution separately in the top panels. Their corresponding relative corrections, $\delta=\sigma_{\text{EW}}/\sigma_{\text{LO}}-1$, are shown in the bottom panels, where we also provide the full NLO EW relative correction (blue). We note that we have investigated the potential impact of using different photon PDFs, namely $\texttt{NNPDF31\_nlo\_as\_0118\_luxqed}$, $\texttt{MMHT2015qed\_nlo}$ \cite{Harland-Lang:2019pla} and $\texttt{CT18qed}$ \cite{Xie:2021equ} sets. At both total and differential cross section levels, we found agreement of the results for photon-induced processes calculated by applying these three photon PDF sets within the statistical uncertainty of the MC integration. The results we show here have been calculated by using the $\texttt{NNPDF31\_nlo\_as\_0118\_luxqed}$ set. In the right-hand-side plot of each figure, we display the LO results (black) for the distribution together with NLO EW (orange) and NLO QCD cross sections with QCD scale uncertainties (light blue band) in the top panel. In the middle panel, the relative NLO QCD corrections, $\delta_{\text{QCD}}=\sigma_{\text{QCD}}/\sigma_{\text{LO}}-1$, calculated by using the central value of factorization and renormalization scale (blue) along with QCD scale uncertainties (light blue band) are shown, where their sizes in percentage are scaled down by a factor of 10 for a better fit in the plot. In the bottom panel, the NLO EW corrections are combined with NLO QCD corrections in the additive (orange) 
\begin{align}
    \sigma_{\text{QCD}\oplus\text{EW}} &=\sigma_{\text{QCD}}+\sigma_{\text{EW}}-\sigma_{\text{LO}}=\sigma_{\text{LO}}(1+\delta_{\text{QCD}}+\delta_{\text{EW}}), \nonumber\\
    \delta_{\text{QCD}\oplus\text{EW}} &=\frac{\sigma_{\text{QCD}\oplus\text{EW}}}{\sigma_{\text{QCD}}}-1=\frac{\delta_{\text{EW}}}{1+\delta_{\text{QCD}}}, 
\label{eq:ew add}
\end{align}
and multiplicative (red) approach
    \begin{align}
    \sigma_{\text{QCD}\otimes\text{EW}} &=\sigma_{\text{LO}}(1+\delta_{\text{QCD}})(1+\delta_{\text{EW}}), \nonumber\\
    \delta_{\text{QCD}\otimes\text{EW}} &=\frac{\sigma_{\text{QCD}\otimes\text{EW}}}{\sigma_{\text{QCD}}}-1=\delta_{\text{EW}},
\label{eq:ew multiply}
\end{align}
where the combined corrections are shown as relative corrections to the NLO QCD results. In this way, we can more clearly assess whether the effects of NLO EW corrections are visible by comparing the combined corrections with the band of NLO QCD scale uncertainties (light blue). 
As can been seen in FIGs.~\ref{fig:m_34_nlo_in} - \ref{fig:pt_z_nlo}, these two approaches of combining NLO EW and NLO QCD corrections cannot be distinguished given the large scale uncertainty of the NLO QCD predictions. This is expected, since the numerical difference is at the level of mixed EW-QCD ${\cal O}(\alpha \alpha_\text{s})$ corrections. It is interesting to note though that these effects are largest in the high  $p_{\text{T},e^+\nu_e}$ region of FIG. \ref{fig:pt_w_nlo} and increase with increasing $p_{\text{T},e^+\nu_e}$. In the following we will discuss the impact of the NLO EW corrections on the distributions in more detail. 

In FIG. \ref{fig:m_34_nlo_in} and FIG. \ref{fig:m_34_nlo} we show the NLO EW and NLO QCD corrections to the invariant mass distribution of the $\mu^+\mu^-$ pair in the low- and high-invariant mass region respectively. In the low-invariant mass region (FIG. \ref{fig:m_34_nlo_in}) the NLO QCD corrections are dominant over the NLO EW corrections. The quark-induced and photon-induced EW corrections are both flat beyond 100 GeV and have opposite signs, yielding an overall relative NLO EW corrections of $-5\%$. There are two peaks at LO, one around 91 GeV is due to the $Z$-boson resonance and the other around 70 GeV comes from the resonance in the $\mu^{+}\mu^{-}\gamma$ three-body invariant mass where the photon is emitted by one of the muons and which results in a shift of the peak in $M_{\mu^+\mu^-}$. 
The distortion of the shape of the $M_{\mu^+\mu^-}$ distribution in the $Z$-boson resonance region is due to collinear final-state photon radiation off muons, which shifts events to smaller values of $M_{\mu^{+}\mu^{-}}$. This distortion is less pronounced when  applying a recombination procedure for muons and photons as done here (see Section~\ref{sec:inputs}). These features in the LO and NLO EW  $M_{\mu^+\mu^-}$ distributions can also be observed in $l^+l^-\gamma$ production as discussed, e.g., in Ref.~\cite{Denner:2015fca}. In the high-invariant mass region (FIG. \ref{fig:m_34_nlo}), the quark-induced EW corrections decrease the LO cross section by $\sim-30\%$ at 1 TeV while the photon-induced corrections are at most $+5\%$, so that the NLO EW corrections go beyond QCD scale uncertainties around 200 GeV, reach $\sim-25\%$ at 1 TeV and become comparable in size to NLO QCD corrections. The NLO EW and NLO QCD corrections have opposite signs and cause a large cancellation in this kinematic region, resulting in a significant change of the shape of the NLO QCD distribution. As discussed earlier, we expect to encounter large negative NLO EW corrections at high-energy tails of distributions for the quark-induced processes due the occurrence of EW Sudakov logarithms.

In FIG. \ref{fig:mt_12_nlo}  we show the NLO EW and NLO QCD corrections to the distribution in the transverse mass of the $e^+ \nu_e$ pair.
The patterns of NLO EW corrections in the transverse mass distribution are very similar to those in the invariant mass distribution of the $\mu^+ \mu^-$ pair. The NLO QCD corrections are uniformly $\sim+30\%$ above about $300$ GeV, while the NLO EW corrections start to become visible beyond the QCD scale uncertainties around 200 GeV and grow negatively in size up to $\sim -25\%$ at 1.5 TeV. The large cancellation between the NLO EW and QCD corrections again significantly changes the shape of the NLO QCD distribution.

In FIG. \ref{fig:m_345_nlo} we present the invariant mass distribution of the $\mu^+\mu^-$ pair and the isolated photon. To this distribution, the NLO QCD corrections are particularly large, ie. $+40\%\sim+80\%$ over the entire invariant mass region. The quark-induced and photon-induced EW corrections show a modest cancellation and yield NLO EW corrections ($\sim -10\%$ at 1 TeV) smaller in size than those in the aforementioned distributions.  The shape of the NLO QCD distribution is barely changed due to the large QCD $K$-factor and scale uncertainties. More MC statistics and even higher-order QCD corrections would be needed to reveal the potential impact of NLO EW corrections.

Besides the invariant mass distributions, we looked into some angular distributions as well. FIG. \ref{fig:r_34_nlo} displays the distribution of angular separation of the $\mu^+\mu^-$ pair, i.e. $R_{ij}$ of Eq.~(\ref{eq: angular separation}) with $i=\mu^+$ and $j=\mu^-$. The photon-induced and quark-induced EW corrections largely cancel when the angular separation is small, i.e. $R_{\mu^{+}\mu^{-}}<2.5$. The quark-induced EW corrections become dominant when the $\mu^+$ and $\mu^-$ are well separated, producing a $\sim -20\%$ overall EW corrections in the region $R_{\mu^{+}\mu^{-}}>3.5$. FIG. \ref{fig:r_15_nlo} shows the distribution of the angular separation
between the positron and photon, i.e.
$R_{ij}$ of Eq.~(\ref{eq: angular separation}) with $i=e^+$ and $j=\gamma$. A similar cancellation happens between the quark- and photon-induced EW corrections at small angular separation, ie. $R_{e^{+}\gamma}<4$, while the photon-induced EW corrections start to be dominant over the vanishing quark-induced ones when $R_{e^{+}\gamma}>4$, resulting into an overall positive EW correction of $\sim +10\%$. However, these subtle patterns of EW corrections appear in the angular region where they are overwhelmed by large NLO QCD scale uncertainties.

\begin{figure}[!tbp]
\centering
\begin{minipage}[b]{0.49\textwidth}
    \includegraphics[width=\textwidth]{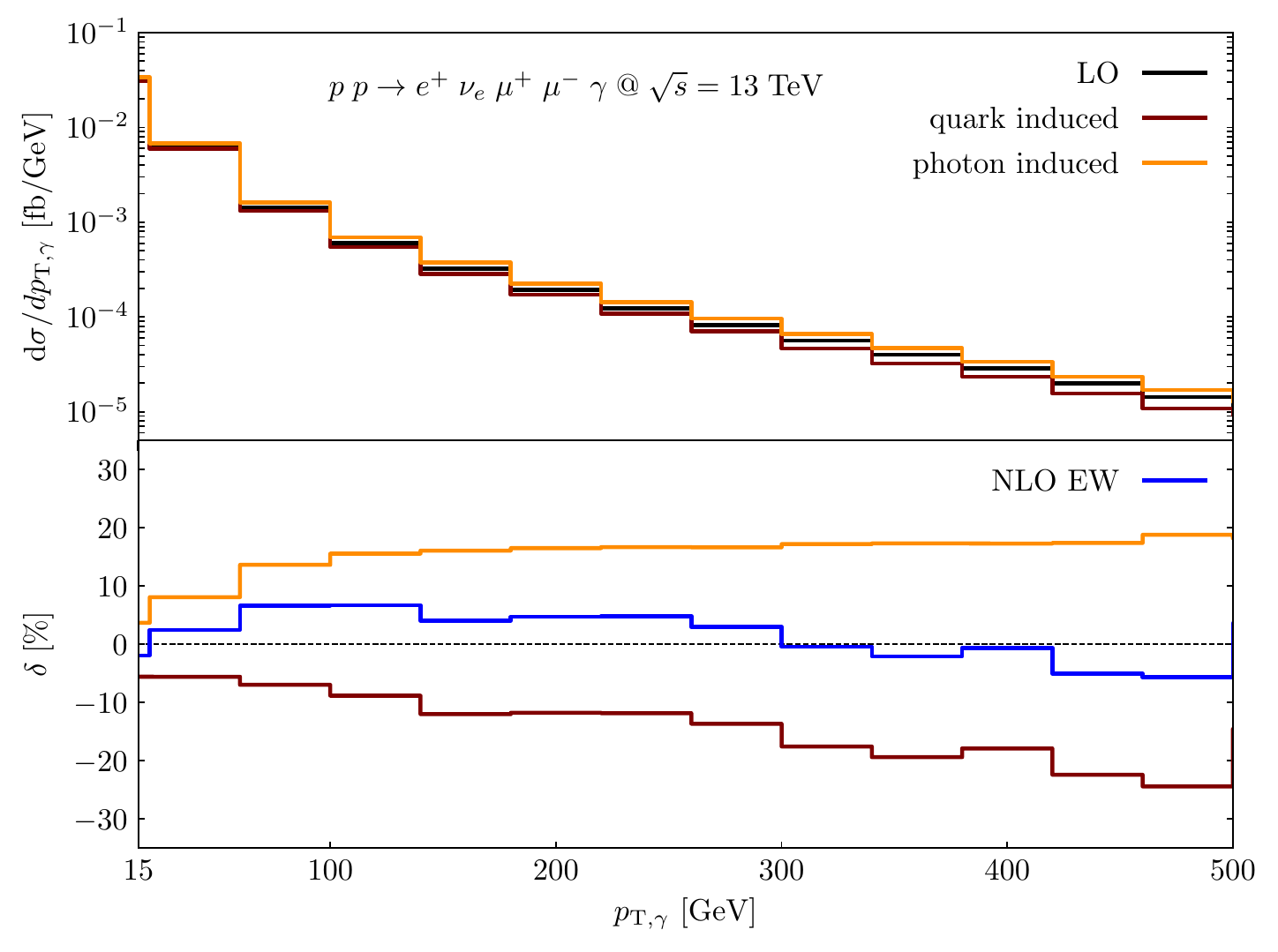}
\end{minipage}
\hfill
\begin{minipage}[b]{0.49\textwidth}
    \includegraphics[width=\textwidth]{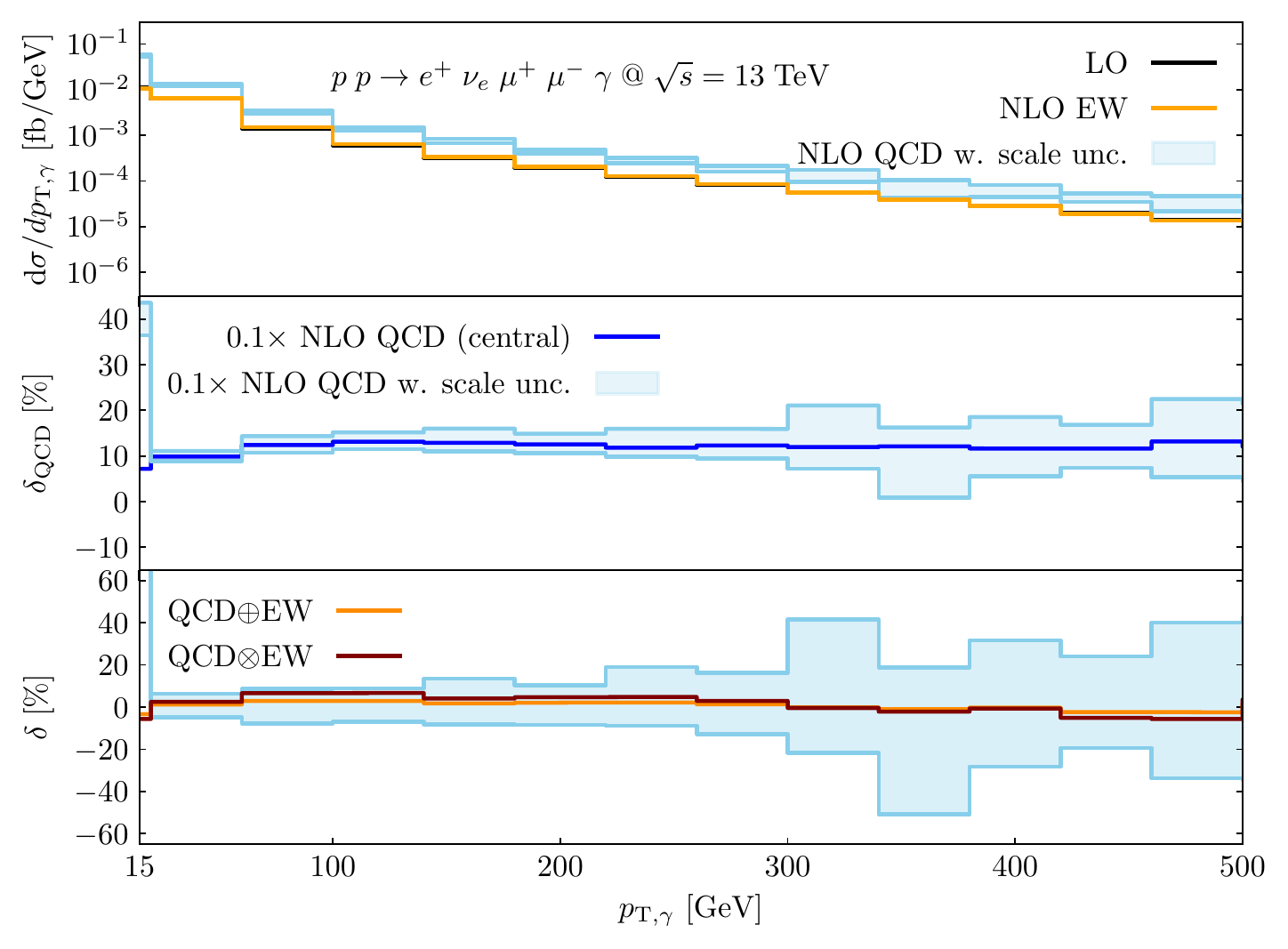}
\end{minipage}
\caption{LO, NLO QCD, and NLO EW predictions for the distribution in the transverse momentum of the photon, and the corresponding relative corrections. See the caption of FIG.~\ref{fig:m_34_nlo_in} and the text for a detailed description.}
\label{fig:pt_5_nlo}
\end{figure}

In FIG. \ref{fig:pt_1_nlo} - FIG.\ref{fig:pt_5_nlo} we present the NLO corrections to various transverse momentum distributions: the $p_\text{T}$ of the $e^{+}$, $e^{+}\nu_e$ pair, $\mu^{+}$, $\mu^{+}\mu^{-}$ pair, the photon and missing transverse momentum. The NLO EW corrections exhibit very similar behaviors among the distributions of the transverse momentum of the positron (FIG. \ref{fig:pt_1_nlo}), missing transverse momentum (FIG. \ref{fig:pt_miss_nlo}) and the transverse momentum of the $e^{+}\nu_e$ system (FIG. \ref{fig:pt_w_nlo}). In these distributions, the photon-induced EW corrections enhance the LO distribution with increasing transverse momentum and cause a $+10\% \sim 20\%$ overall EW corrections at 500 GeV, despite of the typical Sudakov-like negative quark-induced EW corrections. This is because of a new channel firstly opening up at NLO EW where the initial-state photon couples with a virtual $W^{+}$-boson which decays into a positron and an electron neutrino. A related Feynman diagram is shown in FIG.~\ref{fig:ew_real_feynman}. Given such relatively large positive EW $K$-factors, the multiplicative combination of EW and QCD corrections manifests a noticeable difference from the additive one, and tends to break away from the QCD scale uncertainties in the distribution of $p_{\text{T},e^{+}\nu_{e}}$ at 500 GeV. This again reaffirms the necessity of including photon-induced corrections at NLO EW and having a good control of the precision of photon PDFs in such distributions for triboson production processes involving $W$ bosons. The NLO corrections to the transverse momentum distribution of $\mu^{+}$ is displayed in FIG. \ref{fig:pt_3_nlo}. The photon-induced corrections are uniformly about $+10\%$ in the transverse momentum region of $p_{\text{T},\mu^{+}}>100$ GeV and are overwhelmed by the Sudakov-like negative quark-induced EW corrections. The NLO EW corrections reach $\sim -20\%$ at 500 GeV, however, its impact is obscured by large QCD scale uncertainties. FIG. \ref{fig:pt_z_nlo} and FIG. \ref{fig:pt_5_nlo} depict the NLO corrections to the transverse momentum distributions of the $\mu^{+}\mu^{-}$ pair and of the isolated photon respectively. In both distributions, the photon-induced and quark-induced EW corrections have opposite signs and are comparable in size, resulting into small NLO EW corrections ($|\delta_{\text{EW}}|<10\%$) throughout the entire transverse momentum region. Compared to the huge NLO QCD corrections ($\delta_{\text{QCD}}>+100\%$) at 500 GeV, the effects of NLO EW corrections are negligible within QCD scale uncertainties.

\subsection{Effects of dimension-8 operators in SMEFT: An example}
\label{sec:smeft}

\begin{figure}[!tbp]
\centering
\begin{minipage}[b]{0.49\textwidth}
    \includegraphics[width=\textwidth]{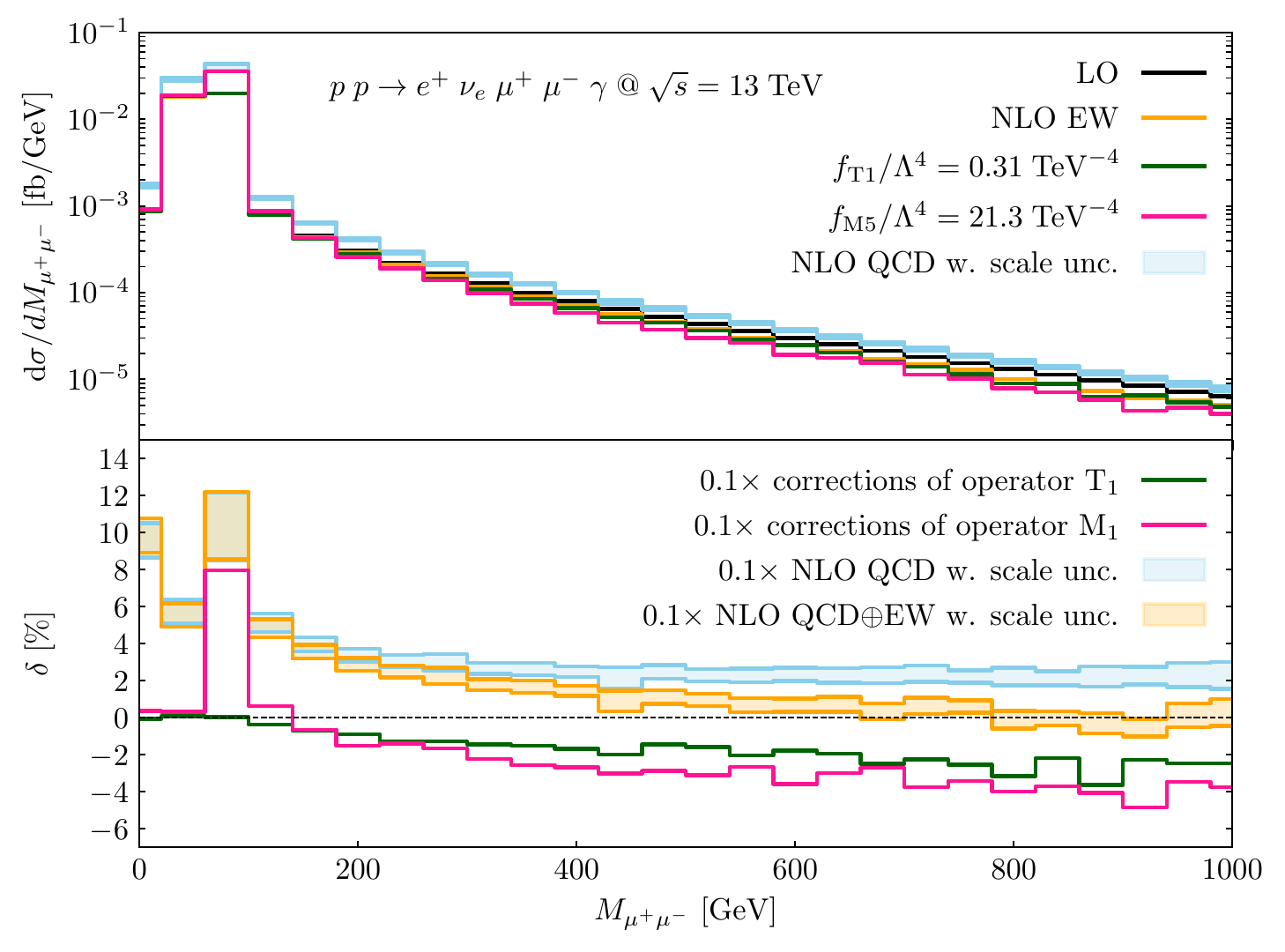}
\end{minipage}
\hfill
\begin{minipage}[b]{0.49\textwidth}
    \includegraphics[width=\textwidth]{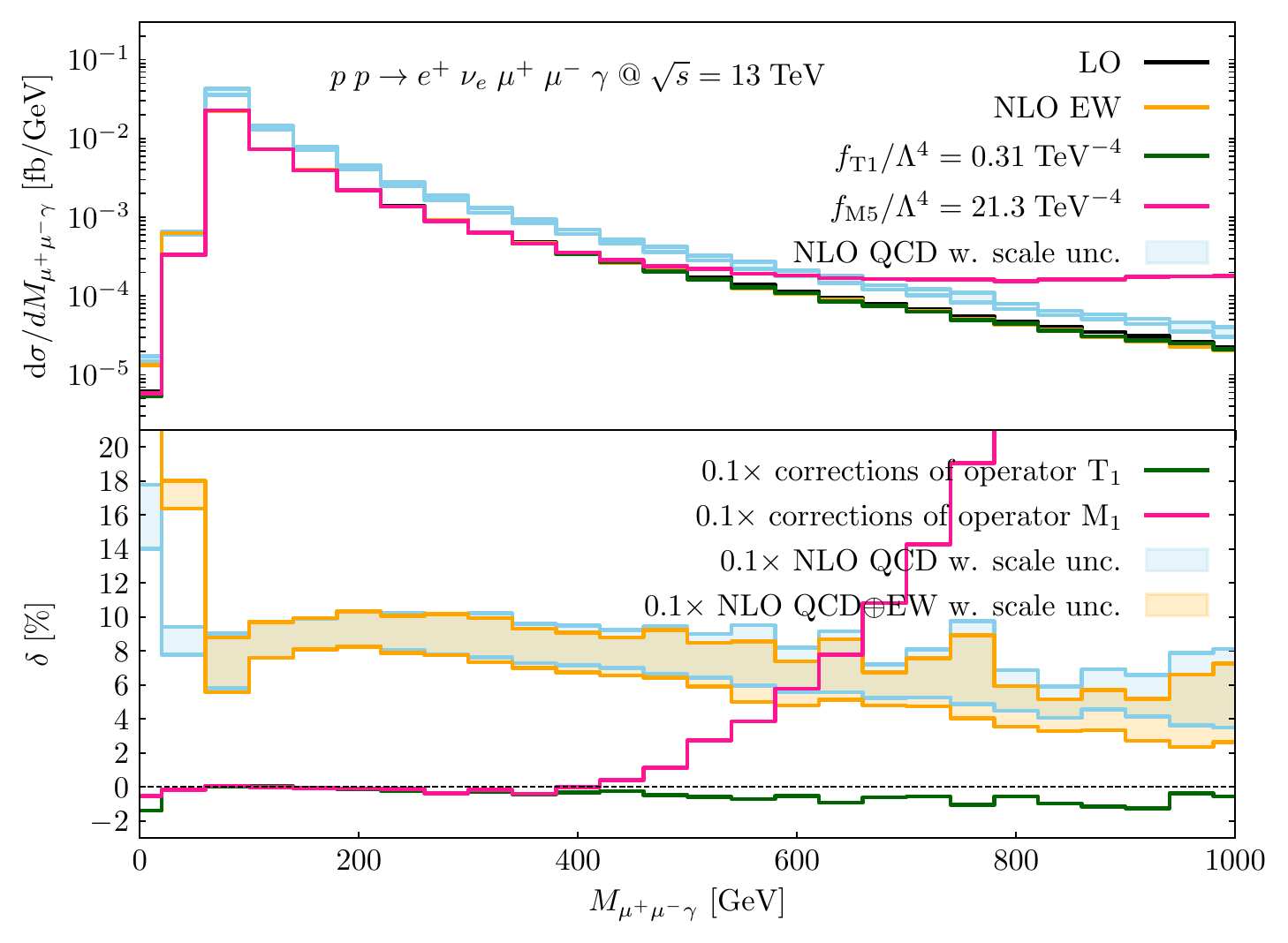}
\end{minipage}
\caption{Predictions for the $\mu^+\mu-$ (left) and $\mu^+\mu^-\gamma$ (right) invariant mass distributions are shown in the top panels at NLO QCD with scale uncertainties (light blue), at NLO+EW QCD (orange) of Eq.~(\ref{eq:ew add}) and at LO including dimension-8 operator $\mathcal{O}_{\text{T}_{1}}$ (green) and $\mathcal{O}_{\text{M}_{5}}$ (pink). The bottom panels show the corresponding relative corrections with respect to the LO prediction.}
\label{fig:eft}
\end{figure}

As we have seen, NLO EW corrections can play a crucial role in kinematic distributions in regions where we expect to be especially sensitive to the effects of BSM physics, namely in the high-energy tails. Therefore, they could also have an impact on the reliability of indirect searches  for BSM physics. To illustrate this point we choose the SMEFT framework~\cite{Brivio:2017vri} and as an example study the impact of dimension-8 operaters. SMEFT obeys SM symmetries and assumes that the new particles are heavy. It is constructed by adding higher dimensional operators to the SM Lagrangian,
\begin{align}
    \mathcal{L}_{\text{SMEFT}}\equiv\mathcal{L}_{\text{SM}}+\sum_{d>4}\sum_{i}\frac{f_{i}^{(d)}}{\Lambda^{d-4}}\mathcal{O}_{i}^{(d)},
\label{eq:smeft}
\end{align}
where $f_i^{(d)}$ is the Wilson coefficient, $\Lambda$ denotes the energy scale of new physics, $d$ is the mass dimension of operators and $\sum_i$ takes the sum over all operators in a given basis for a specific dimension. Here, we consider two dimension-8 operators \cite{Eboli:2006wa} 
\begin{align}
    \mathcal{O}_{\text{M},5}=\left [ (D_{\mu}\Phi)^{\dagger}\hat{W}_{\beta\nu}D^{\nu}\Phi\right]\times B^{\beta\mu},\quad 
    \mathcal{O}_{\text{T},1}=\text{Tr}\left [ \hat{W}_{\alpha\nu}\hat{W}^{\mu\beta} \right ]\times \text{Tr}\left [ \hat{W}_{\mu\beta}\hat{W}^{\alpha\nu} \right ],
\end{align}
individually in Eq.(\ref{eq:smeft}) which only affect the $WWZ\gamma$ and $WW\gamma\gamma$ QGCs in our process. We study the LO effects of a single dimension-8 operator, i.e. the interference between the LO SM and SMEFT amplitudes which contain only one non-standard QGC induced by this single dimension-8 operator, $\mathcal{O}_{\text{M},5}$ or $\mathcal{O}_{\text{T},1}$. We utilize the available UFO model \cite{Degrande:2011ua} for these operators and use $\texttt{MadGraph5\_aMC@NLO}$ to calculate the LO predictions for $p\;p\rightarrow e^{+}\;\nu_{e}\;\mu^{+}\;\mu^{-}\;\gamma$ at $\sqrt{s}=13$ TeV. The coefficients of the operators are taken as the experimentally observed upper limits, $f_{\text{M,5}}/\Lambda^{4}=21.3\;\text{TeV}^{-4}$ and $f_{\text{T,1}}/\Lambda^{4}=0.31\;\text{TeV}^{-4}$ \cite{CMS:2020fqz, CMS:2021gme}. In FIG. \ref{fig:eft}, we show distributions of the invariant mass of the $\mu^+\mu^-$ pair (left) and of the $\mu^+\mu^-\gamma$ system (right) at LO, NLO EW, NLO QCD and when including the $\mathcal{O}_{\text{M},5}$ (pink) and $\mathcal{O}_{\text{T},1}$ (green) operators in LO predictions for these distributions. The NLO QCD corrections are shown with QCD scale uncertainties (light blue) and the additive combination of NLO EW and QCD corrections (orange) are displayed as well. The corrections due to $\mathcal{O}_{\text{T},1}$ and the NLO EW corrections both reduce the LO distributions and are comparable in size in the tail region, i.e. both are around $-20\%$ ($M_{\mu^+\mu^-}$) and $-10\%$ ($M_{\mu^+\mu^-\gamma}$) at invariant masses of 1 TeV. The corrections due of $\mathcal{O}_{\text{M},5}$ behave similar to those due to  $\mathcal{O}_{\text{T},1}$ in the $M_{\mu^+\mu^-}$ distribution, but the $M_{\mu^+\mu^-\gamma}$ distribution is much more sensitive to $\mathcal{O}_{\text{M},5}$ and the relative corrections increase rapidly in higher-invariant mass regions. It should be emphasized that this study is just meant for illustration. A more thorough study would require a wider selection of operators, a study for a range of values for their coefficients (though the experimental constraints are applied), possibly both guided by a UV-completed model. Also, the interplay among various operators as well as among different EFT orders needs to be addressed and eventually a calculation of higher-order corrections in SMEFT may be needed\footnote{See, e.g., the progress made in this direction in case of dimension-6 operators as described for example in Ref.~\cite{Degrande:2020evl}}. Such a dedicated study of $WZ\gamma$ production in SMEFT is however beyond the scope of this paper. Through the illustration in FIG.~\ref{fig:eft}, however, one can already see that missing NLO EW corrections in the SM can mimic effects of single dimension-8 operator in some distributions and kinematic regimes. To take the best advantage of gleaning information about higher-dimensional operators from SMEFT interpretations of LHC measurements, many observables over a wider range of kinematic regimes are needed, and a precise and reliable SMEFT interpretation requires precise SM predictions for all these observables where NLO EW corrections are indispensable.

\section{Summary and conclusions}
\label{sec:summary}
We have calculated the NLO EW and NLO QCD corrections to the process
$p\;p \to e^{+}\;\nu_{e}\;\mu^{+}\;\mu^{-}\;\gamma$ at the 13 TeV LHC. This process includes $W^{+}Z\gamma$ production with leptonic decays ($W^{+} \to e^{+}\;\nu_e$ and $Z \to \mu^+\;\mu^-$), and thus is sensitive to the $WWZ\gamma, WW\gamma\gamma$ QGCs. We provided results for the 
total cross sections and kinematic distributions for a basic set of analysis cuts and studied the impact of these corrections taking into account the theoretical uncertainty due to the factorization and renormalization scale variation at NLO QCD. 
We found that NLO EW corrections are small ($\sim +1\%$) at the total cross section level and negligible in view of large NLO QCD corrections. However, in kinematic distributions their impact can be much more pronounced and visible outside the QCD scale uncertainty bands. For example, in the case of the invariant mass of the $\mu^+\mu-$ pair and the transverse mass of the $e^+\nu_e$ pair, the NLO EW corrections reach $-20\%\sim -40\%$, respectively, in the tail regions where they overtake the NLO QCD corrections and significantly change the shapes of the NLO QCD distributions. In other distributions we studied, the NLO EW corrections are within the NLO QCD scale uncertainties, but they still partially cancel the NLO QCD corrections. A closer look at the NLO EW corrections revealed that the photon-induced contributions largely cancel the quark-induced contributions and even become dominant in some distributions or phase space regions, such as the transverse momentum of the positron, the missing transverse momentum, and the large-angle regions of the angular separation distributions. As an illustration for how missing NLO EW corrections may be mistaken as signals of BSM physics, we studied the LO effects of two dimension-8 operators in SMEFT, $\mathcal{O}_{\text{M},5}$ or $\mathcal{O}_{\text{T},1}$. For their coefficients we chose values inspired by experimental constraints and observed that their impact on certain kinematic distributions can be similar to the one caused by NLO EW corrections to the SM predictions. While this is just a first look, we think it still motivates a comprehensive study of this interplay which is however outside the scope of this paper. To conclude, we hope that this study of NLO EW corrections emphasizes again the importance of including  
these corrections in the interpretation of EW triboson data at the LHC, especially when placing constraints on QGCs and TGCs or dimension-8 operators in SMEFT. Our MC framework, based on $\texttt{RECOLA}$ and $\texttt{MadDipole}$, has been constructed sufficiently flexible, so that it can be readily adjusted to provide LHC predictions for other SM processes at NLO EW and NLO QCD accuracy up to the same level of final-state particle multiplicity.

\section*{Acknowledgements}{We are grateful to Nicolas Greiner for helpful discussions about certain details of the implementation of the dipole subtraction method in $\texttt{MadDipole}$, and for providing us with an updated version of the package as discussed in the text. We thank the University at Buffalo's Center for Computational Research for their support and for providing the necessary computational resources for this work. This work is supported in part by the U.S. National Science Foundation under Grant No. PHY-1719690 and PHY-2014021.}

\newpage
\section*{Appendix}
\subsection{Cancellation of the IR poles at NLO EW at a single phase-space point \label{app:pole}}

We show in Table \ref{table: EW pole cancellations} the cancellation of the IR poles in the case of NLO EW corrections at a single phase-space point which is provided in Table \ref{table: phase space point}.  

\begin{table*}[!ht]
\begin{tabular}{|c|c|c|c|c|}
\hline
        particle & $E$ [GeV] & $p_{x}$ [GeV] & $p_{y}$ [GeV] & $p_{z}$ [GeV] \\
\hline
\hline
        $u$ & 4256.7427754402188 & 0.0000000000000000 & 0.0000000000000000 & 4256.7427754402188 \\
        $\bar{d}$ & 5979.0726006031064 & 0.0000000000000000 & 0.0000000000000000 & -5979.0726006031064 \\
        $e^{+}$ & 2651.1274242259151 & 1139.7908949198131 & -2354.3939455228428 & 431.48868423865929 \\
        $\nu_{e}$ & 3028.0645408716769 & 146.66509096981196 & 2302.0330884744794 & -1961.7104461015310 \\
        $\mu^{+}$ & 1953.8411555790535 & 19.079202693769730 & -1953.5848186647006 & 25.250773094505576 \\
        $\mu^{-}$ & 422.06742678206683 & -398.74993201934308 & 52.968320246807991 & 127.80360525132417 \\
        $\gamma$ & 2180.7148285846151 & -906.78525656405236 & 1952.9773554662563 & -345.16244164584595 \\
\hline
\end{tabular}
\caption{A random phase-space point for the process $u\;\bar{d}\rightarrow e^{+}\;\nu_{e}\;\mu^{+}\;\mu^{-}\;\gamma$ used to check the cancellation of IR poles at NLO EW shown in Table~\ref{table: EW pole cancellations}.} 
\label{table: phase space point}
\end{table*}

\begin{table*}[!ht]
\begin{tabular}{|l|r|r|r|}
\hline
         & \multicolumn{1}{c|}{$1/\epsilon^2$ [$\text{GeV}^{-6}$]} & \multicolumn{1}{c|}{$1/\epsilon$ [$\text{GeV}^{-6}$]} & \multicolumn{1}{c|}{finite [$\text{GeV}^{-6}$]} \\
\hline
\hline
         $\text{EW}_{\text{loop}}$ & $-6.9562240175\times 10^{-28}$ & $-2.0930027056\times 10^{-27}$ & $-1.1458906081\times 10^{-25}$ \\
         $\text{EW}_{\text{I}}$ & $6.9562245027\times 10^{-28}$ & $2.0930028113\times 10^{-27}$ & $-8.1066154115\times 10^{-27}$ \\
\hline
\end{tabular}
\caption{The coefficients of double and single IR poles arising in the interference of the EW virtual amplitude with the LO amplitude ($\text{EW}_{\text{loop}}$) calculated by $\texttt{RECOLA}$ and the integrated dipole $\text{EW}_{\text{I}}$ calculated by $\texttt{MadDipole}$, for $u\;\bar{d}\rightarrow e^{+}\;\nu_{e}\;\mu^{+}\;\mu^{-}\;\gamma$, as well as their finite contributions, at a random phase space point given in Table \ref{table: phase space point}.} 
\label{table: EW pole cancellations}
\end{table*}

\subsection{Some recalculated total cross sections of various processes \label{app:recalculated}}

In this appendix, we present in Table~\ref{table:wwz fixed scale},\ref{table:wwz dynamic scale},\ref{table:e+e-} and \ref{table:e+e-a} the results for NLO QCD and NLO EW contributions to total cross sections for a selection of EW gauge boson production processes at the LHC which are available in the literature and have been recalculated using our in-house MC program.

\begin{table*}[!ht]
\begin{tabular}{|c|c|c|c|c|c|}
\hline
    $\mu_{F}=\mu_{R}=2M_{\text{W}}+M_{\text{Z}}$ & $\sigma_{\text{LO}}$ [fb] & $\sigma_{qg,\bar{q}g}$ [fb] & $\delta_{qg,\bar{q}g}$ [$\%$] & $\sigma_{q\bar{q}}$ [fb] & $\delta_{q\bar{q}}$ [$\%$] \\
\hline
\hline
    In-house MC & 99.23(9) & 49.26(8) & 49.64(9) & 47.9(4) & 48.3(4) \\
    Ref.~\cite{Nhung:2013jta} & 99.29(2) & 49.29(1) & 49.6 & 48.83(3) & 49.2 \\
\hline
\end{tabular}
\caption{LO results, gluon-induced and quark-induced contributions of NLO QCD corrections to $p\;p\to W^{+}\;W^{-}\;Z$ at $\sqrt{s}=14$ TeV, with fixed renormalization and factorization scales, calculated by our in-house MC program and in Ref.~\cite{Nhung:2013jta}.} 
\label{table:wwz fixed scale}
\end{table*}

\begin{table*}[!ht]
\begin{tabular}{|c|c|c|c|c|c|}
\hline
    $\mu_{F}=\mu_{R}=2M_{\text{WWZ}}$ & $\sigma_{\text{LO}}$ [fb] & $\sigma_{qg,\bar{q}g}$ [fb] & $\delta_{qg,\bar{q}g}$ [$\%$] & $\sigma_{q\bar{q}}$ [fb] & $\delta_{q\bar{q}}$ [$\%$] \\
\hline
\hline
    In-house MC & 95.86(9) & 33.99(7) & 35.45(8) & 52.6(4) & 54.9(4) \\
    Ref.~\cite{Nhung:2013jta} & 95.91(2) & 34.07(1) & 35.5 & 53.33(3) & 55.6 \\
\hline
\end{tabular}
\caption{LO results, gluon-induced and quark-induced contributions of NLO QCD corrections to $p\;p\to W^{+}\;W^{-}\;Z$ at $\sqrt{s}=14$ TeV, with dynamic renormalization and factorization scales, calculated by our in-house MC program and in Ref.~\cite{Nhung:2013jta}.} 
\label{table:wwz dynamic scale}
\end{table*}

\begin{table*}[!ht]
\begin{tabular}{|c|c|c|c|}
\hline
    $M_{ll}>50$ GeV & $\sigma_{0}$ [pb] & $\delta_{q\bar{q},\text{weak}}$ [$\%$] & $\delta^{\text{rec}}_{q\bar{q},\text{phot}}$ [$\%$] \\
\hline
\hline
    In-house MC & 738.7(4) & -0.719(4) & -1.81(1) \\
    Ref.~\cite{Dittmaier:2009cr} & 738.733(6) & -0.71 & -1.81 \\
\hline
\end{tabular}
\caption{LO results, weak corrections and photonic corrections to $p\;p\to l^{+}\;l^{-}+X$ at $\sqrt{s}=14$ TeV, with $M_{ll}>50$ GeV, calculated by our in-house MC program and in Ref.~\cite{Dittmaier:2009cr}.} 
\label{table:e+e-}
\end{table*}

\begin{table*}[!ht]
\begin{tabular}{|c|c|c|c|c|c|}
\hline
    $\sqrt{s}=14$ TeV & $\sigma^{\text{LO}}$ [pb] & $\delta_{\text{weak},q\bar{q}}$ [$\%$] & $\delta^{\text{CS}}_{\text{phot},q\bar{q}}$ [$\%$] & $\delta_{\gamma\gamma}$ [$\%$] & $\delta^{\text{Frix}}_{\text{QCD}}$ [$\%$]\\
\hline
\hline
    In-house MC & 1317.1(9) & -0.76(1) & -2.96(5) & 0.2229(7) & 67.7(5) \\
    Ref.~\cite{Denner:2015fca} & 1317.4(1) & -0.74 & -2.70(1) & 0.22 & 67.09(7) \\
\hline
\end{tabular}
\caption{LO results, weak corrections, photonic corrections, photon-initiated contributions and NLO QCD corrections with the Frixione isolation cut applied, to $p\;p\to e^{+}\;e^{-}\;\gamma$ at $\sqrt{s}=14$ TeV, calculated by our in-house MC program and in Ref.~\cite{Denner:2015fca}.} 
\label{table:e+e-a}
\end{table*}

\newpage
\bibliography{references}

\end{document}